\RequirePackage{fix-cm}
\documentclass[twocolumn]{svjour3}          
\smartqed  
\usepackage{graphicx}
\usepackage{natbib}
\usepackage{amsmath}
\usepackage[table]{xcolor}
\usepackage[dvipdfm]{hyperref}
\hypersetup{
   colorlinks=true,             
   linkcolor=blue,              
   urlcolor=blue,               
   anchorcolor=blue,            
   citecolor=blue,              
}

\definecolor{grey80}{rgb}{0.90,0.90,0.90}

\journalname{Celestial Mechanics and Dynamical Astronomy}
\begin{document}

\title{Machine Learning applied to asteroid dynamics}

   \author{V. Carruba$^{1}$  \and
           S. Aljbaae$^{2}$  \and
           R. C. Domingos$^{3}$  \and
           M. Huaman$^{4}$  \and
           W. Barletta$^{1}$  \and
   }
   
   \authorrunning{Carruba et al. 2021} 
      \institute{V. Carruba \at
     \email{\href{valerio.carruba@unesp.br}{valerio.carruba@unesp.br}} \\
     Orcid ID: 0000-0003-2786-0740 \\
      \and
      \begin{itemize}
        \item [$^{1}$] S\~{a}o Paulo State University (UNESP), School of Natural Sciences and Engineering, Guaratinguet\'{a}, SP, 12516-410, Brazil
        \item [$^{2}$] National Space Research Institute (INPE), Division of Space Mechanics and Control, C.P. 515, 12227-310, S\~{a}o Jos\'e dos Campos, SP, Brazil
        \item [$^{3}$] S\~{a}o Paulo State University (UNESP), Sao Jo\~{a}o da Boa Vista, SP, 13876-750, Brazil
        \item [$^{4}$] Universidad tecnol\'{o}gica del Per\'{u} (UTP), Cercado de Lima, 15046, Per\'{u}  
      \end{itemize}
   }
   \date{Received: date / Accepted: date}   
   \maketitle

\begin{abstract}
  Machine Learning ($ML$) is the branch of computer
  science that studies computer algorithms that can learn from data.
  It is mainly divided into supervised learning,
  where the computer is presented with examples of entries, and the goal is
  to learn a general rule that maps inputs to outputs, and unsupervised
  learning, where no label is provided to the learning algorithm, leaving
  it alone to find structures. Deep learning is a branch of machine learning
  based on numerous layers of artificial neural networks, which
  are computing systems inspired
  by the biological neural networks that constitute animal brains. In
  asteroid dynamics, machine learning methods
  have been recently used to identify members of asteroid families, small bodies
  images in astronomical fields, and to identify resonant arguments images of
  asteroids in three-body resonances, among other applications.
  Here, we will conduct a full review of available literature in the field, and
  classify it in terms of metrics recently used by other authors to assess the
  state of the art of applications of machine learning in other
  astronomical subfields.  For comparison, applications
  of machine learning to Solar System bodies, a larger
  area that includes imaging and
  spectrophotometry of small bodies, have already reached a state classified as
  {\it progressing}.  Research communities and
  methodologies are more established,
  and the use of $ML$ led to the discovery of new celestial objects
  or features, or new insights in the area. $ML$ applied to asteroid
  dynamics, however, is still in the {\it emerging} phase, with smaller
  groups, methodologies still not well-established, and fewer papers
  producing discoveries or insights.
  Large observational surveys, like those conducted at the
  Zwicky Transient Facility or at the Vera C. Rubin Observatory,
  will produce in the next years very substantial datasets of orbital
  and physical properties for asteroids. Applications of $ML$ for
  clustering, image identification, and anomaly detection, among others,
  are currently being developed and are expected of being of great
  help in the next few years.
   
\end{abstract}

\section{Introduction}
\label{sec: intro}

Astronomical datasets are rapidly growing in size and complexity, ushering
in the era of big data science in Astronomy
\citep{2010IJMPD..19.1049B, 2010AdAst2010E..58P}.
As discussed in the recent review paper by \citet{2019arXiv190407248B},
the sheer magnitude of recent datasets in astronomy demands the introduction
of techniques other than the visual inspection by a human researcher.
Clustering methods to identify groups of objects sharing similar
properties, like galaxies, asteroids, or variable stars, can now
be performed with much simpler to write software, yielding faster
results. New possible members of known groups in large databases can 
be automatically assigned to their most likely class.
Anomaly detection algorithms can now identify outliers
and new classes of astronomical objects with minimal training from
a human observer.
To stand up to the challenges of modern astronomy, many new techniques
in Machine Learning ($ML$) have been recently adopted by more and more
subfields in astronomy.

Machine learning is a subfield of Engineering and Computer Science.
Machine learning explores the study and construction of
algorithms that can learn from data.  Algorithms for machine
learning are often split into two main categories.
Supervised machine learning algorithms are used to learn a mapping from
a set of features to a target variable based on provided examples of
input-output pairs.  Unsupervised learning algorithms are utilized to learn 
relationships that exist in the dataset without using labels.

\citet{2020WDMKD..10.1349F} recently reviewed the then-cur\-rent status
of applications of $ML$ to various subfields of astronomy, like
{\it Planetary studies, Active Galactic nuclei, Solar Astronomy} and
{\it Variable Stars}, among others.  Applying a set of metrics to
the scientific outcome of the available literature, the authors
were able to classify fields for applications of $ML$
into {\it emerging}, {\it progressing}, and {\it established}.
Fields like {\it Solar Astronomy} and {\it Variable Stars} were
classified as {\it established} by \citet{2020WDMKD..10.1349F},
with large groups operating in the field, a substantial body of literature
available, and scientific outcomes reaching the more advanced phases
of {\it discovery} and {\it insight}, where new scientific knowledge is
demonstrated as a consequence of applying $ML$.

While applications of machine learning in other astronomical subfields
have recently flourished, the number of papers that applied such methods
in asteroids dynamics has been more limited, with just 12 papers that
tried to use such approaches in the last three years. 
The main goal of this paper is to review recent applications
of $ML$ to the field of asteroid dynamics, and to use
\citet{2020WDMKD..10.1349F} metrics to assess the impact
that the use of existing algorithms has had in the field.  Once this
task is done, we will then assess open problems and some possible
future lines of research.

Many of the revised papers used $ML$ algorithms
developed
in the {\it Python} language \citep{CS-R9526}, or,
less frequently, in $R$ \citep{R_ref} or {\it Julia} \citep{bezanson2017julia}.
Here, we will cover three main areas of applications
of $ML$ to asteroid dynamics, mostly associated with applications
of three popular {\it Python} libraries:  i) Supervised and
unsupervised learning, {\it scikit-learn}
\citep{2012arXiv1201.0490P}, ii) time series analysis, {\it statsmodels}
\citep{seabold2010statsmodels} and iii)  deep learning, {\it Tensorflow}
\citep{tensorflow2015-whitepaper} and {\it Keras} \citep{Chollet_2018}.

The most commonly used approaches in asteroid dynamics and
application to small bodies will be covered
in the next sections.  Please note that the purpose of these
sections is not to provide a full description of all available algorithms
in the literature, and of the theory behind them, but rather to offer a
review of the field and to provide resources where interested readers
could find an additional material.  Interested readers can find a more
broad review of the statistical methods discussed in this work in
\citet{2019arXiv190407248B}.  We will start our review by discussing
the most commonly used methods in $ML$.
Less popular techniques, like time series analysis, 
will be discussed in section~(\ref{sec: TS analysis}).  Finally,
applications of deep learning and the use of {\it Artificial Neural Networks}
will be covered in section~(\ref{sec: Deep learning}).  After a full
review of existing literature in the field has been performed, we will
apply the metrics used by \citep{2020WDMKD..10.1349F} to assess 
the impact, use, and current level of diffusion of existing algorithms
in the field in section~(\ref{sec: class_insight}).
Section~(\ref{sec: future trends})
will cover what we consider some likely future trends in the field.
Finally, in section~(\ref{sec: concl}) we will present our conclusions.

\section{Machine Learning}
\label{sec: ML}

Machine learning approaches used in applications to small bodies
are divided into two main categories:

\begin{enumerate}
\item Supervised learning: ``The computer is presented with the task
  of learning a function that maps an input to an output based
  on example input-output pairs'' \citep{russel2010}.
\item Unsupervised learning: no label is given to the learning algorithm,
  leaving the method alone to find the structure at its entrance
  \citep{Wang_2001}.
\end{enumerate}

Another category of $ML$, reinforcement learning, has not yet
had applications in the area of interest of this work and will not be
further discussed.
$ML$ algorithms can either predict a label for  
{\it classification} problems, or a quantity, and in this case,
one has {\it regression} problems.
We will start our review by looking into supervised training algorithms.

\subsection{Supervised learning}
\label{sec: sup_learn}

Concerning supervised learning, several algorithms have been
developed in the {\it Python} language in recent years.
They are freely available to the scientific community
({\it scikit-learn} package, \citep{2012arXiv1201.0490P}).
Supervised learning is defined by the use of labeled datasets to
train algorithms that classify data or predict outcomes accurately.
Supervised learning algorithms can either be standalone or ensemble methods.

Standalone methods are approaches based on a single algorithm.  Some of the most
used standalone methods in astronomy are {\it Linear regression},
{\it Logistic regression}, and {\it Decision tree}.

\begin{figure}
  \includegraphics[width=3.5in]{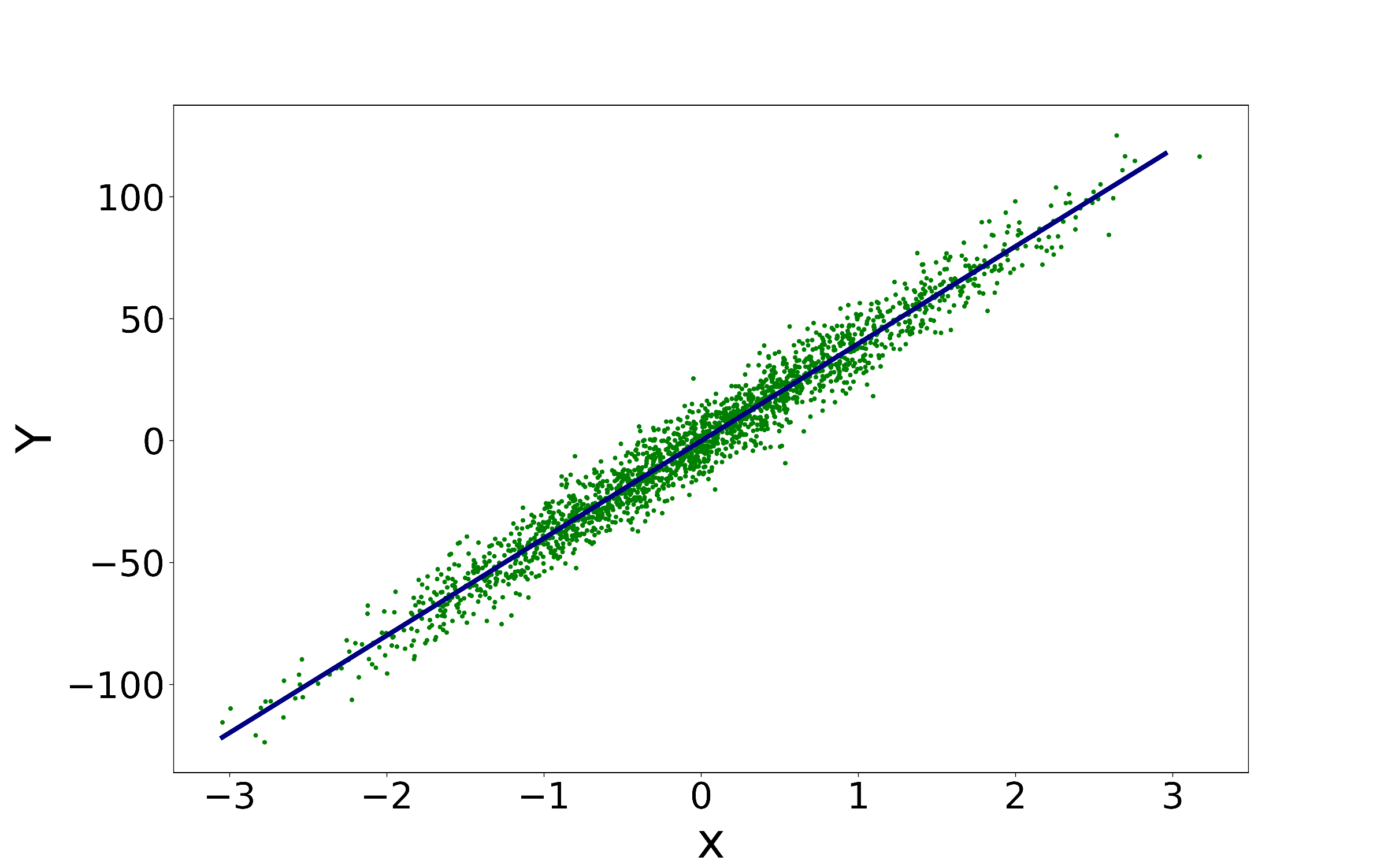}
  \caption{An application of linear regression to randomly generated
    data.  The data was created using the {\it make\_regreession} function
    from the libray {\it sklearn.datasets}.}
\label{Fig: linear_regression}
\end{figure}

{\it Linear regression} looks
for a linear relationship between an independent variable, {\it x}
or the input, and a dependent variable, {\it y} or the
output. The algorithm fits a linear model to minimize the residual sum
of squares difference between observed and predicted data, as shown
in figure~(\ref{Fig: linear_regression}). In this method,
it is possible to force the coefficients to be positive, which could be
useful for some physical problems.

\begin{figure}
  \includegraphics[width=3.5in]{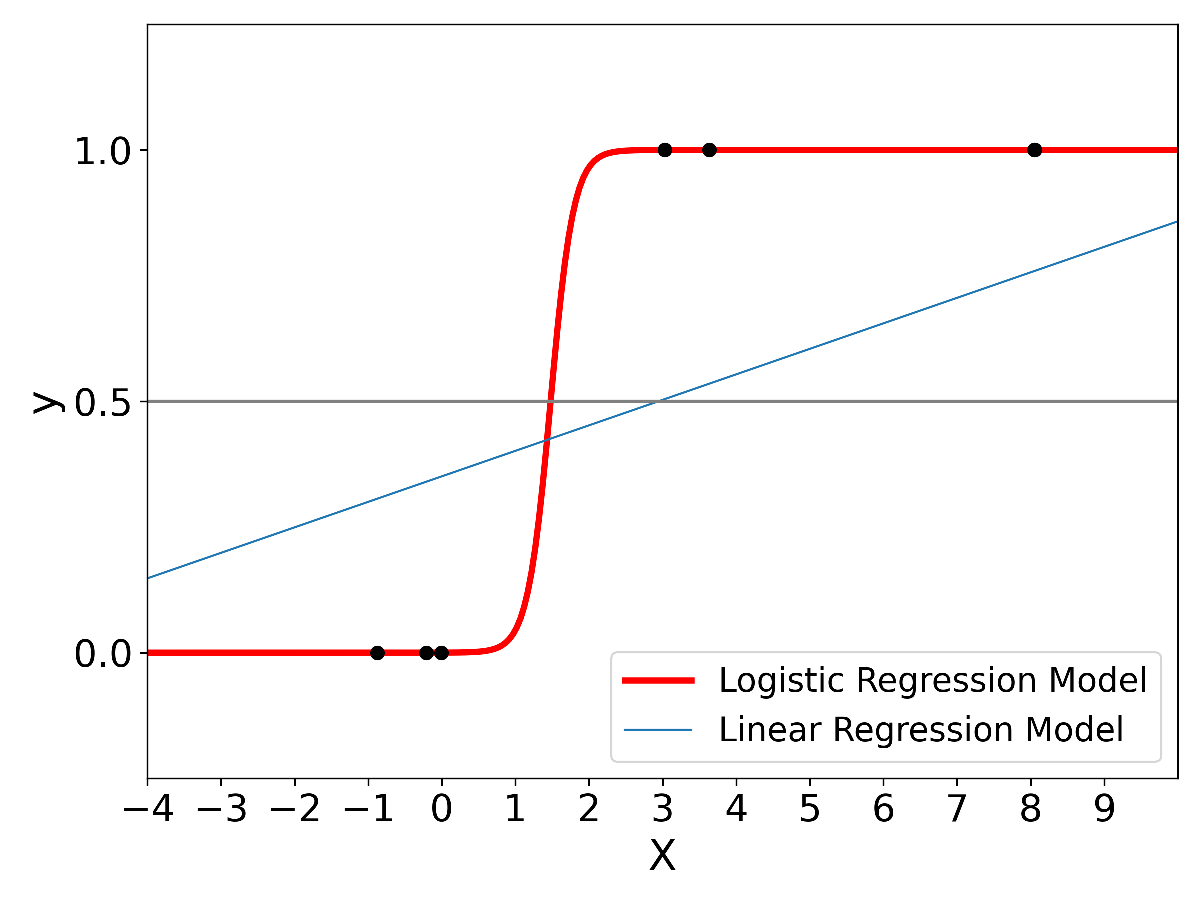}
  \caption{An example of {\it logit} function, superimposed with case
    of linear regression.}
\label{Fig: logistic_regression}
\end{figure}

{\it Logistic regression} is a statistical method applied to
cases where the input data needs to be labeled into
different categories \citep{Dalpiaz_2021}. It is a special case
of linear regression
where the output variable is a single value. {\it Logistic
regression} uses probability theory to predict whether
anything belongs to a group or not. This type of problem
is known as a “binary classification problem” as one
classifies an object as either belonging to a group or not.
In this case, there are two possible results: “yes” and
“no”. Usually, a positive result points to the presence of
some identity separate from those of its members while a
negative result points to the absence of it. Thus, one
needs to predict the probability that identity is present.
In practice, the binary responses variable is coded using 1
and 0 for “yes” and “no”, respectively.

To do so, we fit the function {\it logit} to the set of training data to map
any real value to value in the interval between 0 and 1 based on a
linear combination of dependent predictor variables as:

\begin{equation}
  logit{P_i}=log\left(\frac{P_i}{1-P_i}\right)={\beta}_0+{\beta}_1 x_{1,i}
  +{\beta}_2 x_{2,i}+...+{\beta}_k x_{k,i},
\label{eq: logistic_regr}
\end{equation}

where $P_i$ is the probability of the event i term, and
the coefficients ${\beta}_i (i = 1, 2, ..., k)$ are parameters of the
model usually estimated through the maximum likelihood estimation
(MLE) and $x_i (i = 1, 2, ..., k)$ is a vector
containing discrete or continuous values
\citep{CRAMER2004613, GUDIVADA2016169}.  An example of {\it logit}
function is shown in figure~(\ref{Fig: logistic_regression}).

\begin{figure}
  \includegraphics[width=3.5in]{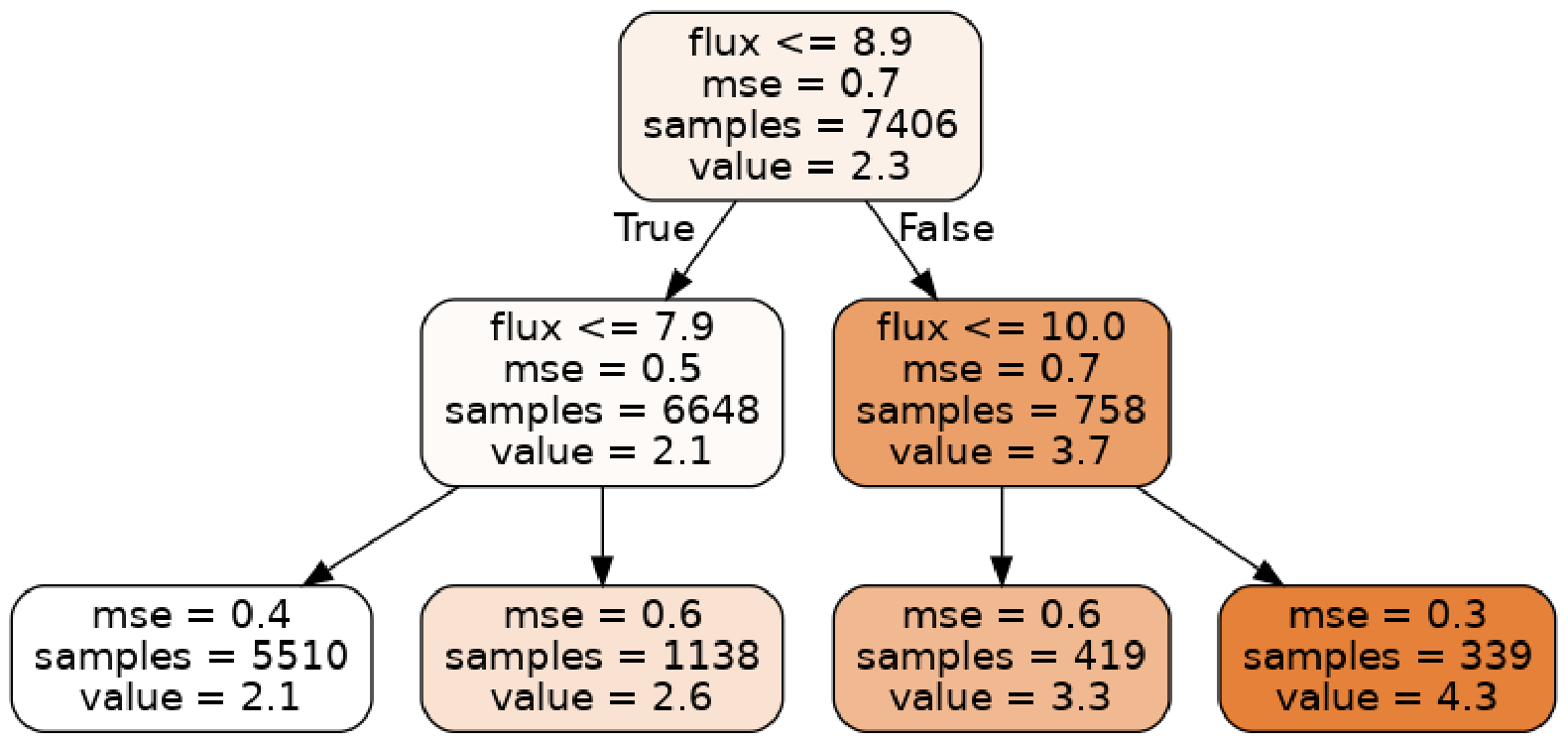}
  \caption{Decision tree regressor. 7406 objects from the Gaia Data Release 2
    \citep{2018A&A...616A..13G} are used to train a decision tree algorithm to
    found correlations between the relative flux uncertainties and the diameter
    measurements of the object \citep{2021RNAAS...5..199D}. }
\label{Fig: dec_tree}
\end{figure}

A {\it Decision tree} is a model that is described by a top-to-bottom
tree-like graph.  It can be used in both classification and regression.
{\it Decision tree} algorithms make decisions in a tree-shaped
form, as consecutive nodes.  The lowest nodes in the tree are called
leaves or terminal nodes.  They are not associated with a condition, but
instead, carry the label of a path within the tree.  The number of
decision nodes, or the maximum depth, “{\it max$\_$depth}”, is a free
parameter of the model. An example of
a tree diagram for a {\it Decision tree} is shown in
figure (\ref{Fig: dec_tree}), where we display the path to classify
a sample of asteroids from the Gaia Data Release 2 \citep{2018A&A...616A..13G},
as used by \citep{2021RNAAS...5..199D}.
In the first node of this regressor, all the objects with relative flux
uncertainties $\le 9$, in a logarithmic scale, will follow the arrow
marked as True, and the rest will follow the False arrow.  Other decisions
will follow as marked.  

Ensemble methods use several standalone algorithms to obtain a result.
We can distinguish between:

\begin{enumerate}
\item {\it Bootstrap aggregation (bagging)}: the training data is divided into
  several samples, called bootstrap samples, obtained by selecting a random
  sample with replacement of the training set. Each model is counted
  with the same weight.
\item {\it Boosting:} emphasizes the training cases where the previous models
made fewer mistakes in the classification, and the autonomous models that
had a better performance.
\end{enumerate}

\begin{figure}
  \includegraphics[width=0.45\textwidth]{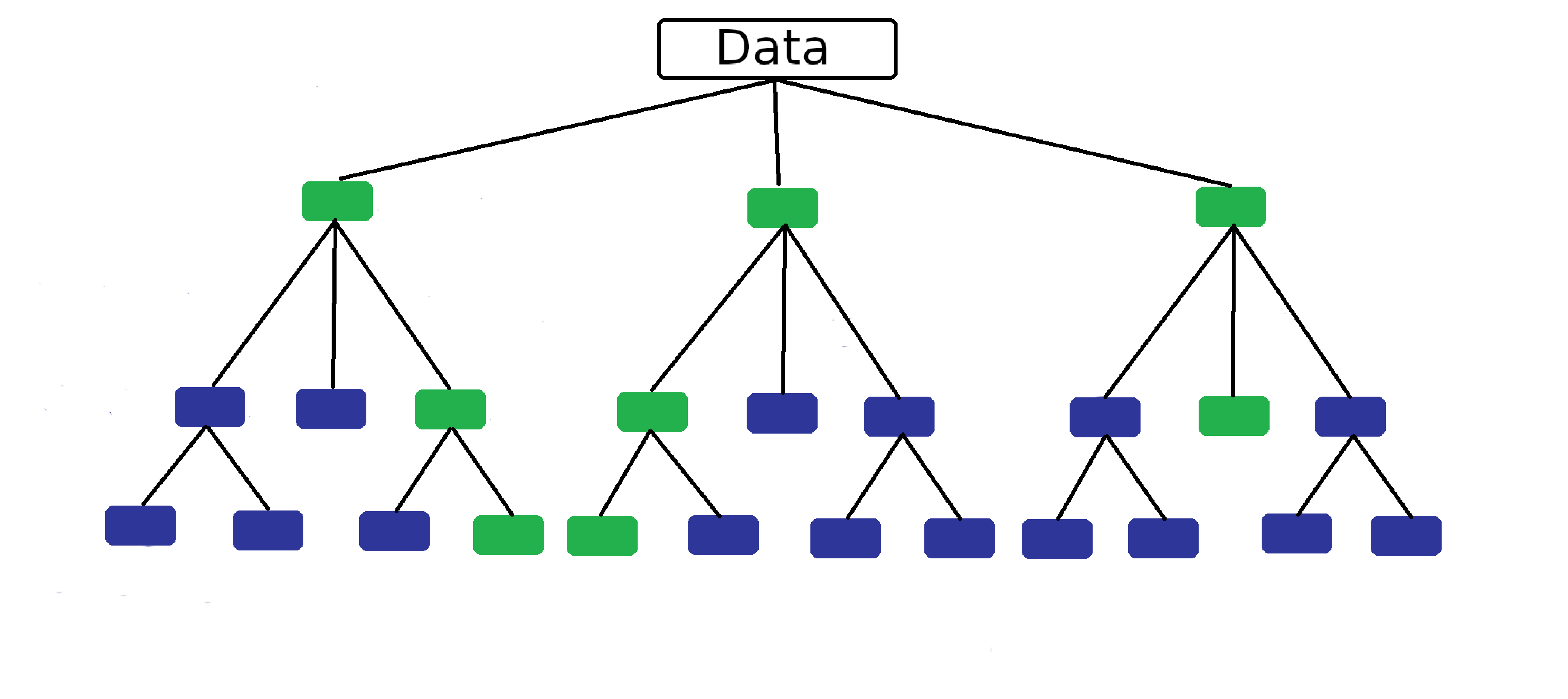}
  \caption{An example of the procedure used by {\it Random Forest} to
    reach decisions.}
\label{Fig: Random_forest}
\end{figure}

Examples of {\it Bagging Classifiers} are the {\it Random Forest} and the
{\it Extremely Randomized Trees}.  For {\it Random Forest}, 
a multitude of decision trees is created at
training time. For classification tasks, the {\it Random forest's} output is
the class chosen by the majority of trees \citep{Ho95,Ho98}.
The mean or average prediction
of the individual trees is returned for regression tasks. 
{\it Extremely Randomized Trees}, or {\it ExtraTree}, follows a similar
procedure, but the bagging process is performed without substitutions.
An example of the procedure used by  {\it Random Forest} to reach an
outcome is shown in figure~(\ref{Fig: Random_forest}).  

\begin{figure*}
  \includegraphics[width=6.in]{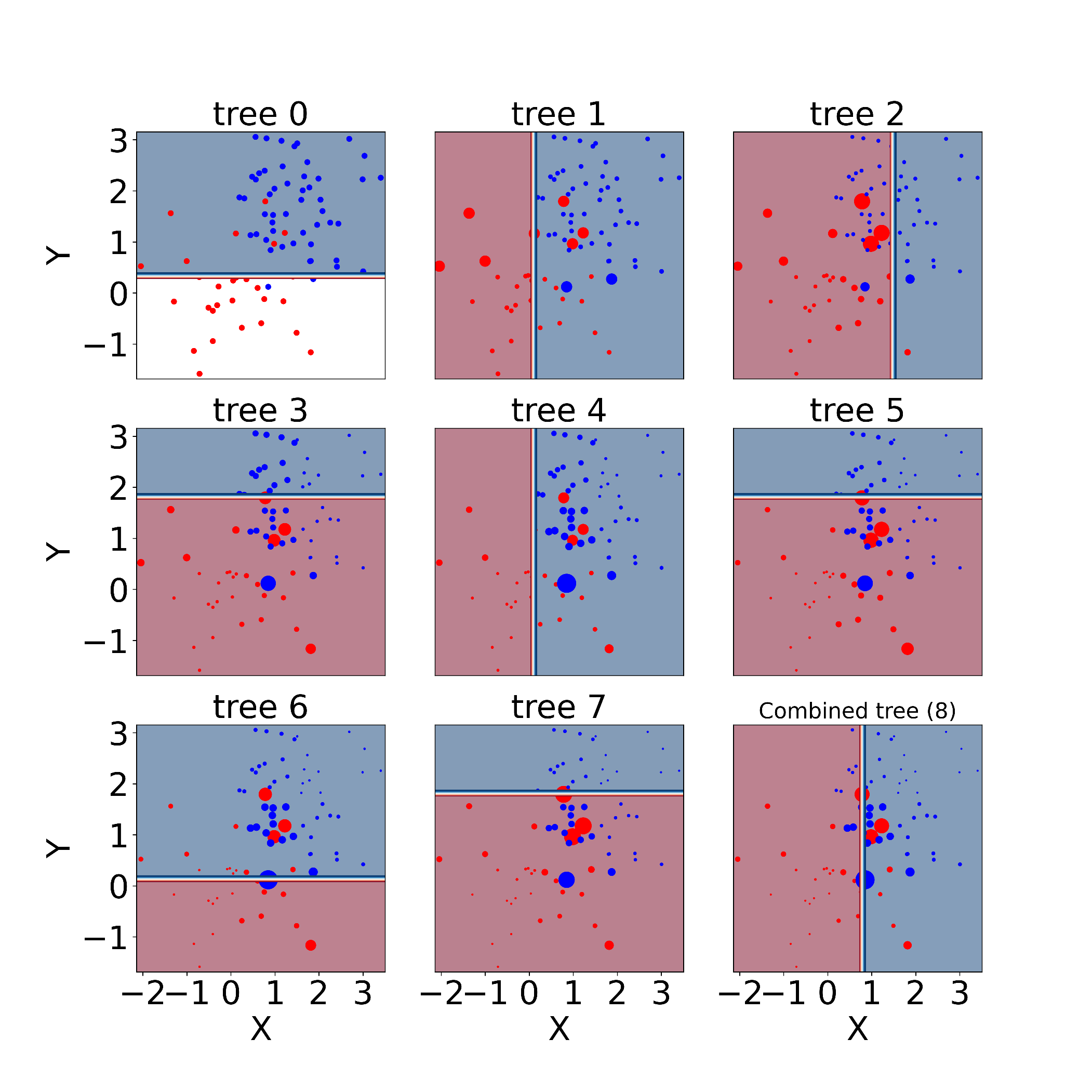}
  \caption{An illustration of the method by which {\it Adaboost} reaches
    decisions. The algorithm combines weak learners, assigns weights to
    incorrect values, and implements sequential tree growing considering
    past mistakes.}
\label{Fig: Adaboost}
\end{figure*}

Examples of  {\it Boosting} algorithms are {\it Adaptive Boosting (AdaBoost)},
{\it Gradient Boosting (Gboost)}, and {\it eXtreme Gradient Boosting (XGBoost)}.
In the {\it AdaBoost} method \citep{Freund95adecision-theoretic}, the
reinforcement algorithms track the models
that provided the least accurate forecast and give them lesser weights.
The final result is achieved as a weighted average of the result of each
autonomous model, as also shown in figure~(\ref{Fig: Adaboost}).
{\it GBoost} differs from {\it AdaBoost} because it employs a
gradient descent algorithm in the process of assigning weights to the
standalone models' outcome \citep{Boehmke19}.  The Gradient algorithm
was regularized and optimized in {\it XGBoost} by \citet{Chen_2016}.

Another example of ensemble tree-based methods are the
{\it Bayesian additive regression trees} ($BART$)
\citep{Chipman_2010, doi:10.1146/annurev-statistics-031219-041110}.
$BART$ is a model in which each tree is limited by a regularization before
being a weak learner, and fitting and inference are carried out using an
iterative Bayesian back-fitting algorithm that generates samples a
posterior. While not a part of the standard {\it Python scikit-learn} package,
$BART$ is available in Python at the {\it Git\-Hub} repository
``https://github.\-com/JakeColtman/bartpy'', last accessed on June 1st 2022.

{\it Naive Bayes} is a simple approach for creating classifiers, which are
models that offer class labels to problem cases represented as vectors of
feature values, with the class labels selected from a limited set.
There is no single technique for training such classifiers; rather,
there are several algorithms based on the same principle: all
{\it Naive Bayes} classifiers assume that the value of one feature is
independent of the value of any other feature, given the class variable.
For example, a fruit is called an apple if it is red, round, and has a
diameter of roughly 10 cm. A robust {\it Naive Bayes} classifier considers
each of the color, roundness, and diameter features to contribute
independently to the likelihood that this fruit is an apple
\citep{Piryonesi_20}, regardless of any possible correlations between
them. The continuous values associated with each class are assumed to
follow a normal distribution in the {\it Gaussian Naive Bayes} method. 

Finally, {\it Support-Vector Machines (SVM)} have also been recently used
for classification and regression problems \citep{Cortes_09}. Assume that
some data points are assigned to one of two classes, and the purpose is to
determine which class a new data point will be assigned to. A data point is
seen as a $p$-dimensional vector (a list of $p$ numbers) in support-vector
machines, and we want to know if we can separate such points with a
$(p-1)$-dimensional hyperplane. Numerous hyperplanes might
be used to categorize the data. The hyperplane that represents the greatest
separation, or margin, between the two classes is a viable choice as the
best hyperplane.  The algorithm upon which {\it SVM}s are based automatically
selects the best hyperplane. 

\subsection{Unsupervised learning}
\label{sec: unsup_learn}

Unsupervised learning is a general term that incorporates a large set of
statistical tools used to perform data exploration.  Here we are going
to review mostly algorithms used for {\it clustering}, which
is the problem of arranging a set of items so that objects in the same group
(called a cluster) are more comparable (in some sense) to those in
other groups (clusters), and for {\it anomaly detection}, which is
a step in data mining that identifies data points, events, or observations that
deviate from a dataset most observed behavior. 
Concerning {\it clustering}, {\it K-Nearest-Neighbors (KNN)},
{\it k-means}, and the {\it Hierarchical Clustering
  Method (HCM)} were the most used methods in recent literature,
while recently published papers that worked with
{\it anomaly detection} used {\it Isolation forests}.
All the algorithms that we outline in this section are available 
in the {\it Python} library scikit-learn \citep{2012arXiv1201.0490P}.

\begin{figure}
  \includegraphics[width=3.0in]{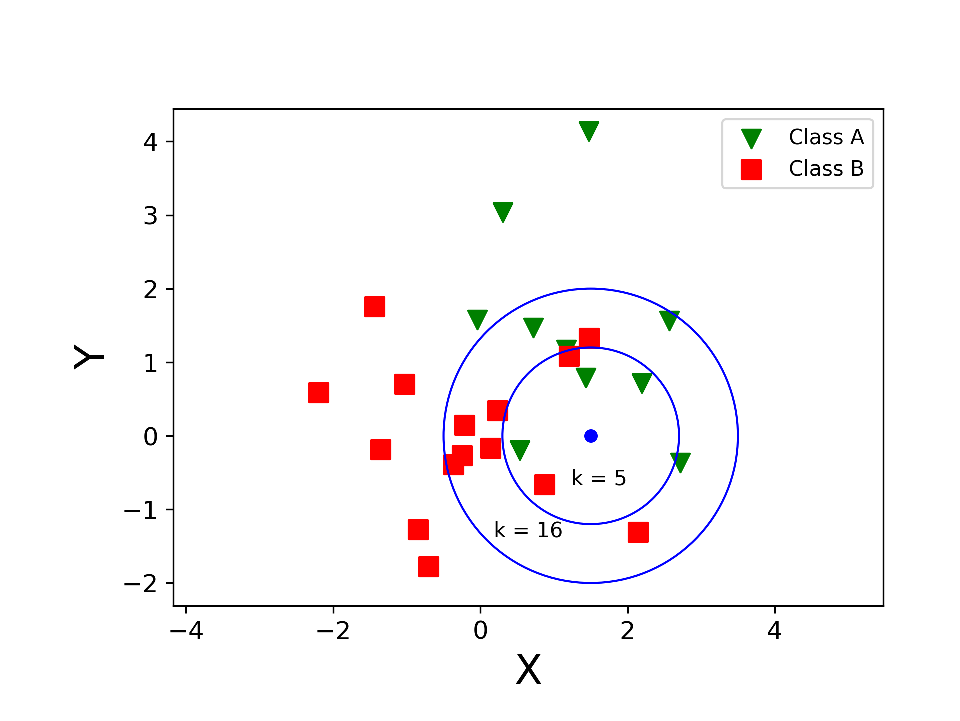}
  \caption{An example of bi-dimensional application of {\it KNN} for
    classification into two classes.}
\label{Fig: KNN}
\end{figure}

The basic idea behind {\it KNN} is to predict the status of
a point based on its closest neighbors.  
If most of the closest neighbors
belong to a class, the point is also classified as such.
A free parameter of this algorithm is the number of neighbors used
for the classification.
Figure~(\ref{Fig: KNN}) shows a case where, if
we use three as a number of neighbors, the central point will be classified
as belonging to class B, while for a larger sample of 7 the point is classified
as belonging to class A. {\it KNN} can be used for both supervised
and unsupervised learning.  Most recent applications used for problems
of unsupervised learning, which is why we decided to list it in this
section.

{\it K-means}, a centroid-based clustering technique, is one of the most
extensively used clustering methods \citep{MacQueen1967}.
The distance assignment between the objects in the sample is the first stage
in {\it K-means}. The Euclidean metric is the default distance, however
other metrics that are better relevant for the dataset at hand can be utilized.
The algorithm then chooses $k$ random items from the dataset to act as the
initial centroids, with $k$ being an external free parameter.
The closest of the $k$ centroids is then allocated to each object in
the dataset. The average position of the objects associated with the
specified cluster is then used to construct new cluster centroids. 
These two procedures, re-assigning objects to clusters based on their
distance from the centroid and recomputing cluster centroids, are performed
until convergence is achieved. Convergence can be described in a variety of
ways, such as when the vast majority of objects (90 percent or more) are no
longer reassigned to various centroids, or when the cluster centroids
converge to a single position. The cluster centroids and an association of
the individual objects to the different clusters are the algorithm's output. 

\begin{figure}
  \includegraphics[width=3.5in]{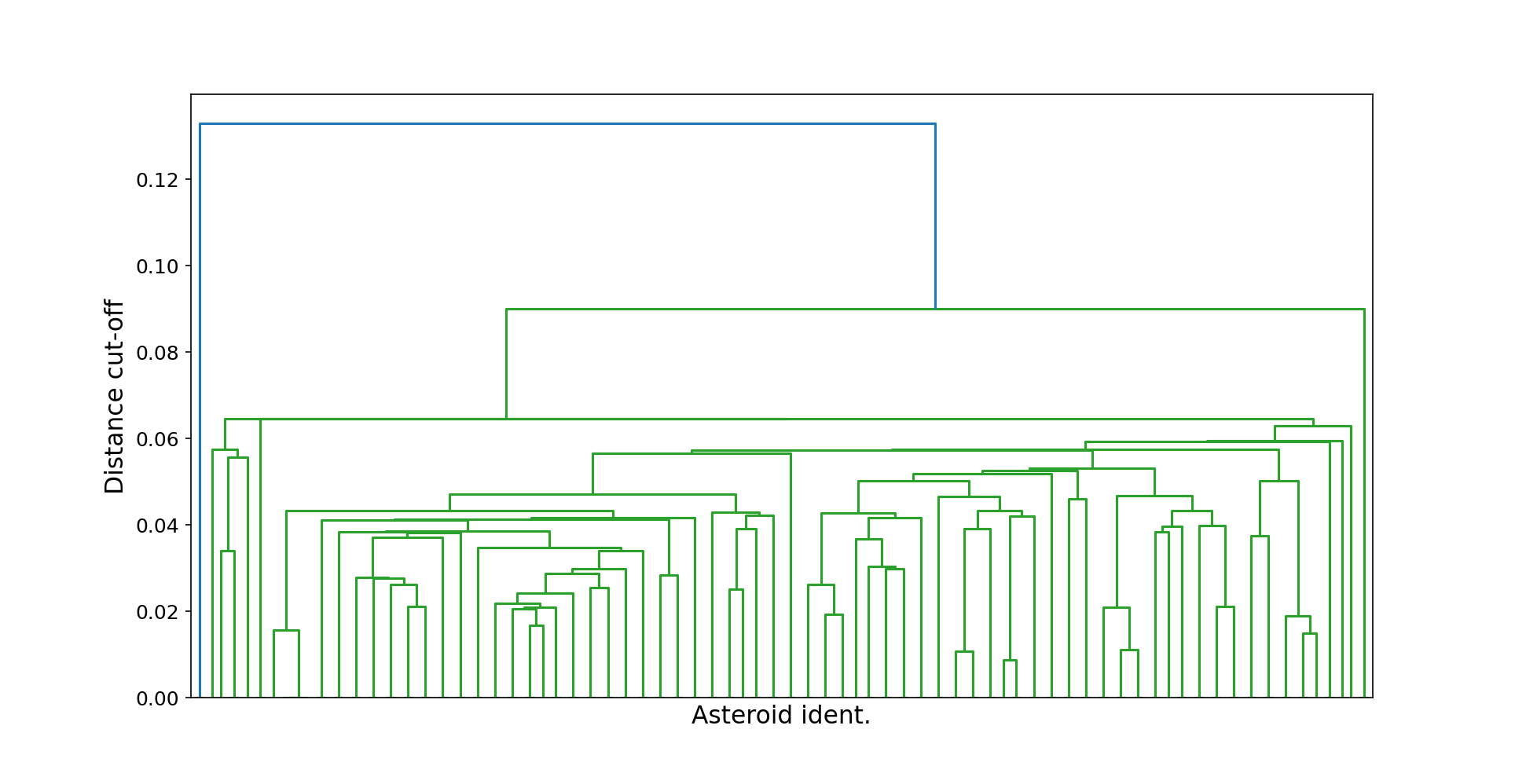}
  \caption{A dendrogram for classifying asteroid lightcurves using $HCM$ in
    domains of three non-periodic quantities.  Adapted from 
    figure~(12) of \citet{Carruba_2021_l}.}
\label{Fig: HCM}
\end{figure}

{\it HCM} is the most commonly used method for identifying
families of asteroids \citep{2002aste.book..613B}. First,
it checks for objects closer to a parent body than to a distance cut-off.
If a new object is found, the new body is used as the
parent body and the process is repeated until
no new members are identified.  Results of {\it HCM} are usually displayed
in form of a dendrogram, as shown in figure~(\ref{Fig: HCM}), where {\it HCM}
has been used to find groups in a domain of three non-periodic quantities
of asteroid light curves.  Each vertical line in the dendrogram represents
a cluster, while each horizontal line is associated with the merging
of groups.  The vertical axis displays the distance cut-off.

Finally, concerning anomaly detection algorithms, {\it Isolation forest}
is the first anomaly detection algorithm to use isolation to identify
anomalies.  It was first conceived and developed by \citet{Liu_08}.
Anomalies are described as those instances in the dataset that do not
conform to the normal profile, according to the most common methodologies
for anomaly detection.  {\it Isolation forest} takes a different approach:
instead of attempting to construct a model of normal examples, it separates
anomalous points in the dataset explicitly. The key benefit of this approach
is the ability to use sampling techniques in ways that profile-based methods
cannot, resulting in a fast algorithm with low memory requirements.

In the next sub-section we are going to review $ML$ methods recently
applied in the fields of time-series analysis.
    
\subsection{$ML$ application in time-series analysis}
\label{sec: TS analysis}

Time-series analysis studies data points collected over a period of time. 
This form of study, on the other hand, is more than just
gathering data over time. The ability to depict how variables change over
time distinguishes time-series data from other types of data.
In other words, time is an important variable since it reveals
how the data changes through time as well as the final outcomes. 

In this review paper, we are going to discuss two $ML$ methods that have been
recently used for studies of asteroid dynamics, $ARIMA$ models and
the use of auto-correlation functions $ACF$.  Other methods, like
the use of Exponential smoothing models ($ETS$) or the use of deep
learning approaches based on artificial neural networks will
not be treated in this section.  Interested readers can find further
details on these topics in \citet{brownlee_2020}.   Numerical methods
discussed in this section are available in the {\it Python} library
{\it statsmodels} \citep{seabold2010statsmodels}.

One exciting area where $ML$ has been recently applied is time-series analysis.
Auto-re\-gressive ({\it AR}), integrated ({\it I}), moving average
({\it MA}), or ARIMA models have been developed to predict stock
prices behavior \citep{RePEc:elg:eebook:1506}.  Recent applications of
$ARIMA$ models include Fraud detection in Credit Card transactions
\citep{2020arXiv200907578M}, studies of variable stars light curves 
\citep{2018FrP.....6...80F}, and studies of coronal index solar cycles
\citep{2020A&C....3200403A}, among others.
  
In an auto-regressive (AR) process current values depend on past ones through
a relationship:

\begin{equation}
  x_t = a_1 x_{t-1}+a_2 x_{t-2} + ...+ a_p x_{t-p} +{\epsilon}_t,
\label{eq: AR}  
\end{equation}

\noindent where ${\epsilon}_t$ is a normally distributed error with
constant variance and zero mean, $p$ is the process order (i.e., how
many time lags are employed in the model), and $a_i (i = 1, 2,...,p)$ are
the model coefficients for each lag up to order $p$. Current values in a
moving average (MA) process are influenced by recent errors, or shocks,
via the following relationship: 

\begin{equation}
  x_t={\epsilon}_t +b_1 {\epsilon}_{t-1} +b_2 {\epsilon}_{t-2}+...+ b_q{\epsilon}_{t-q},
  \label{eq: MA}
\end{equation}

\noindent where ${\epsilon}_t$ is the error term for the {\it t-th} term, and
the coefficients $b_i (i = 1, 2,...,q)$ quantify the response to the previous
shocks up to order $q$.  Adding the two equations produces an
{\it ARMA(p,q)} model.  The coefficients $a_i$ and $b_i$ are estimated using
regression procedures, like the maximum likelihood estimation.

The {\it ARMA} model operates under several assumptions, that include the series
being stationary, i.e., with constant mean and variance. A stationary
series presents the same behavior at all times.  Stationarity in a time series
can be checked by performing tests such as the Dickey-Fuller, also known as
Adfuller test \citep{doi:10.1080/01621459.1979.10482531}.  This test can be
used by verifying the null hypothesis that a unit root is present in an AR
model.  If the probability $p_{AF}$ associated with the null hypothesis is
less than 0.05, the series can be
assumed to be stationary. A non-stationary series can be transformed into a
stationary one by taking the difference of all elements with respect to
their previous ones.
If the new difference series is stationary, a modelization of this new series
can then be carried out.  If that is not the case, a new difference series
can be computed from the order 1 difference series, and then checked
for stationarity.  The process can be then repeated until stationarity
is achieved.

{\it ARIMA} models depend on three free parameters, $(p,d,q)$.
These are the order $p$ of the autoregressive part of the model, the order
$d$ of series differentiation by which stationarity is achieved, and
the order $q$ of the moving average part. These coefficients can be
determined using the Box-Jenkins approach \citep{1976tsaf.conf.....B}.

Other astronomical applications based on time series analysis were based
on the autocorrelation function. Chaos in dynamical systems can be identified
using an approach based on the autocorrelation function of time series, the
ACF index ({\it ACFI}). The Pearson correlation coefficient
\citep{1895RSPS...58..240P} measures how strong two time series are related.
$R$ is near to 1, the maximum value, if the two variables depicted by the
series are tightly connected. $R$ will be close to the minimal value of -1
if they are significantly anti-correlated, and $R \simeq 0$ if there is no
correlation at all. A correlation coefficient can be defined in a variety
of ways, but Pearson's approach is the most frequent. If the i-th term of
the series in $x$ and $y$ is defined as $x_i$ and $y_i$, then: 

\begin{equation}
  R = \frac{cov{(X,Y)}}{{\sigma}_X{\sigma}_Y},
  \label{eq: Pearson_R}
\end{equation} 

\noindent where $cov{(X,Y)}$, which is the covariance of the two series, 
is defined as:

\begin{equation}
  cov{(X,Y)}=\frac{1}{N^2}\sum_{i=1}^{N}\sum_{j=1}^{N}\frac{1}{2}(x_i-x_j)(y_i-y_j).
\label{eq: Covariance}
\end{equation}

\noindent $N$ is the number of terms in the two series, and ${\sigma}_X$
is the standard deviation of the $x_i$ series, defined as:

\begin{equation}
{\sigma}_X = \sqrt{\frac{1}{N}\sum_{i=1}^{N}(x_i-{\mu}_x)^2},
\label{eq: standard_dev}
\end{equation}

\begin{figure*}
  \includegraphics[width=0.49\textwidth]{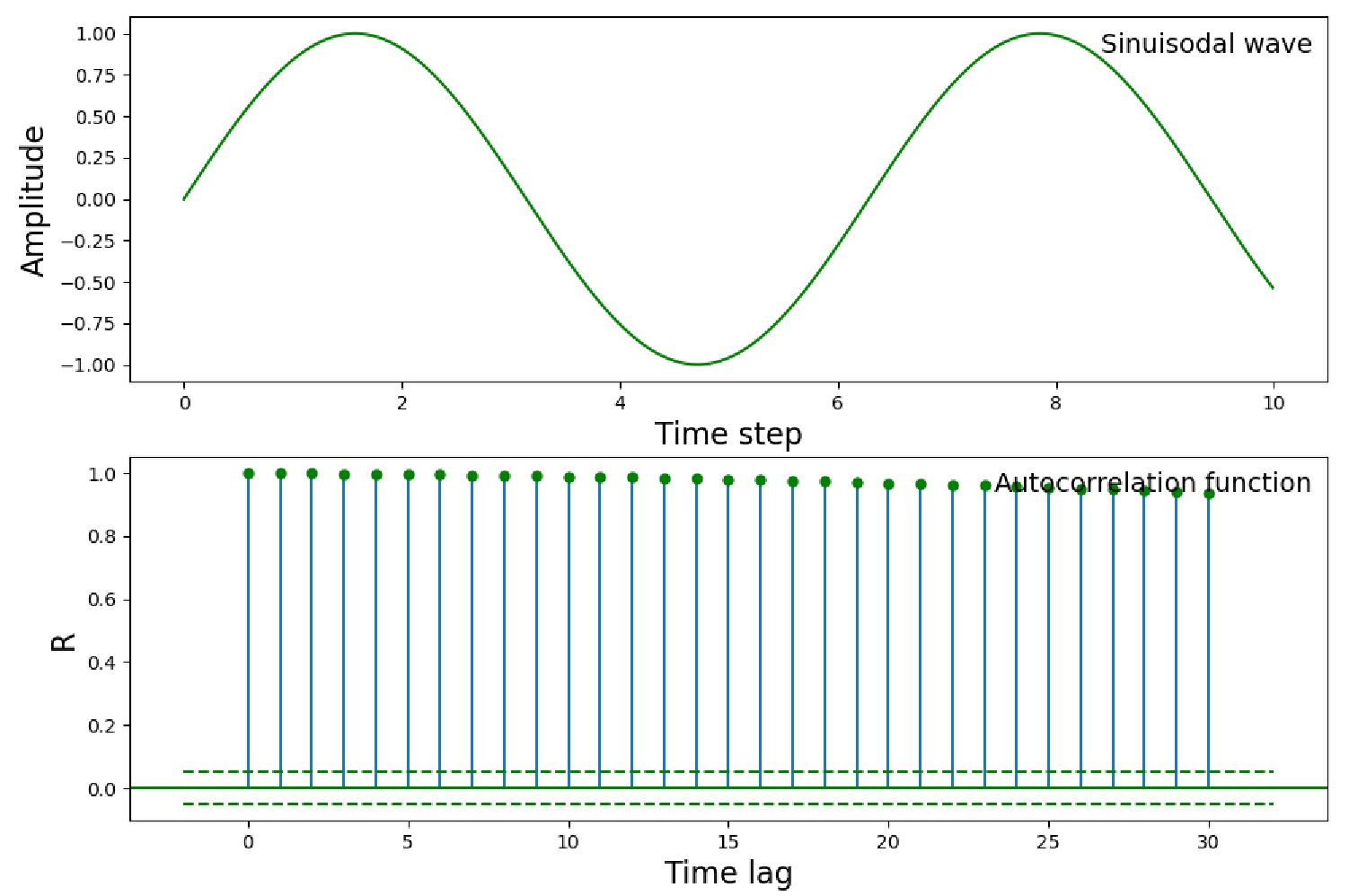}
  \includegraphics[width=0.49\textwidth]{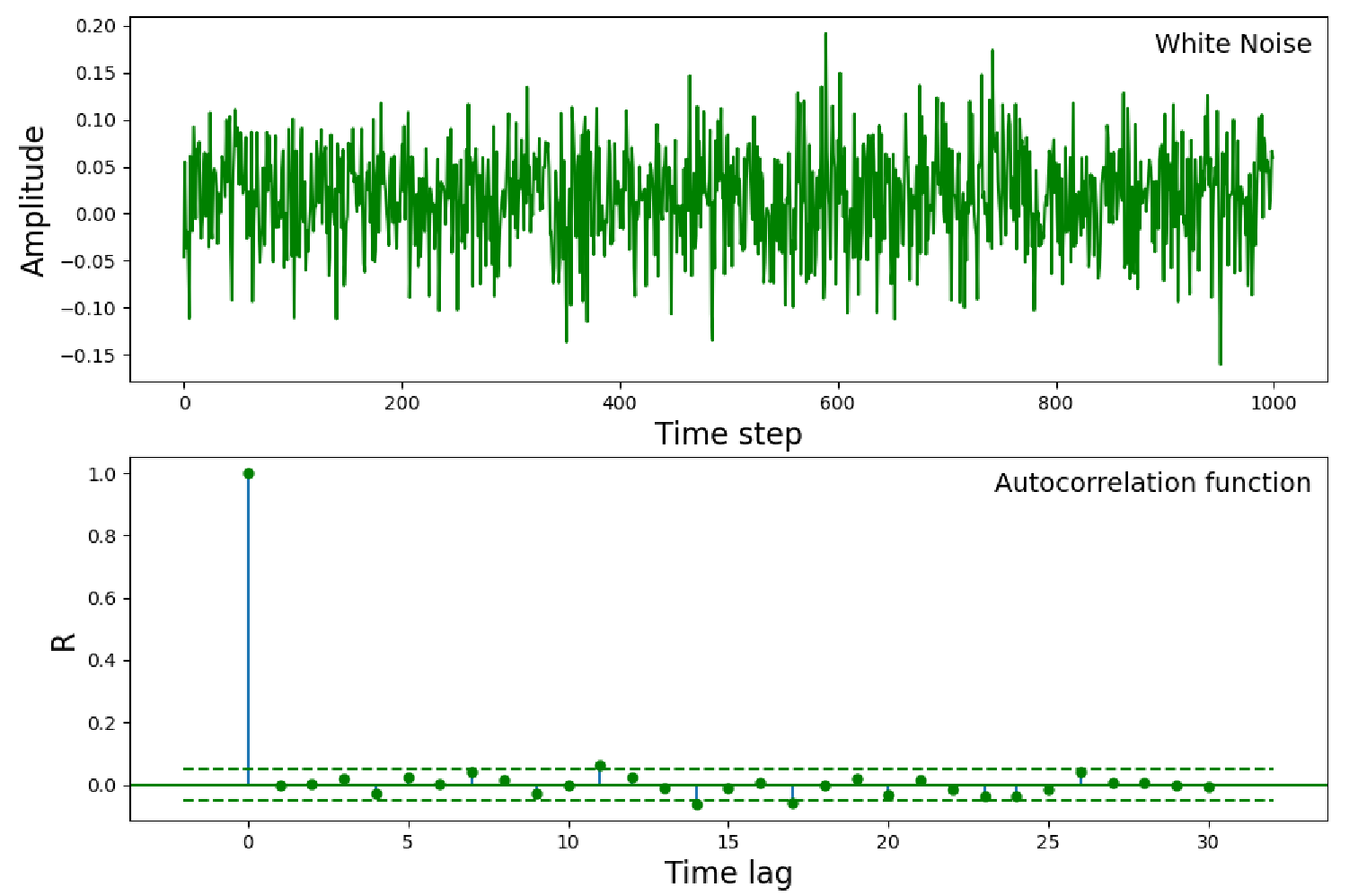}
  \caption{On the left, we have a plot of a sinusoidal wave's time behavior
    (top panel) and its autocorrelation function (bottom panel).
    We have simulated white noise (top panel) and its $ACF$ on the right side
    (bottom panel). In the $ACF$ plots, the area between dashed horizontal
    lines represents the region where autocorrelation coefficients are less
    than 5\% and represent negligible autocorrelation.
    Adapted from figure~(1) of \citep{2021CeMDA.133...24C}.}
\label{Fig: acf_examples}
\end{figure*}

\noindent where ${\mu}_x = \frac{1}{N}\sum_{i=1}^{N} x_i$ is the average value of
the series. For ${\sigma}_Y$, a similar expression exists. The correlation
function of a time series with a lagged copy of the series is the
autocorrelation coefficient of the series. Assume we have built a time
series with a lag of one, $y_i = x_{i-1}$. Equation~(\ref{eq: Pearson_R})
will be used to calculate the autocorrelation coefficient for this $y_i$.
For lags of 2 ($y_i =x_{i-2}$), 3 ($y_i =x_{i-3}$), and so on,
analogous autocorrelation coefficients can be found.
The spectrum of autocorrelation coefficients for various values of
the time lag is the autocorrelation function ($ACF$) of $x_i$. 
The autocorrelation function is useful for determining the predictability
of a series' temporal behavior.

For a function that periodically repeats itself, all its auto-correlation
coefficients are equal to 1 (see figure~(\ref{Fig: acf_examples})). It
is always possible to predict its future behavior.  For white noise time
series, such as random variables, all its auto-correlation coefficients are
around 0. It is not possible to predict the series future behavior. 

We can define a chaos indicator using the ACF, the auto-
correlation function indicator ({\it ACFI}), as:

\begin{equation}
ACFI = \frac{1}{i_{fin}-i_{in}} \sum_{i=i_{in}}^{i=i_{fin}} n_i(\left|R\right|)>0.05),
\label{eq: ACFI}
\end{equation}

\noindent where $n_i(\left|R\right|)>0.05)_i$ is the number of
autocorrelation coefficients larger, in absolute value, than 5\%, the null
hypothesis for auto-correlation. {\it ACFI} is computed over the interval
$i_{fin}-i_{init}$ to avoid including auto-correlation at short timescales.
 
\subsection{Deep learning}
\label{sec: Deep learning}

\begin{figure}
  \includegraphics[width=3.5in]{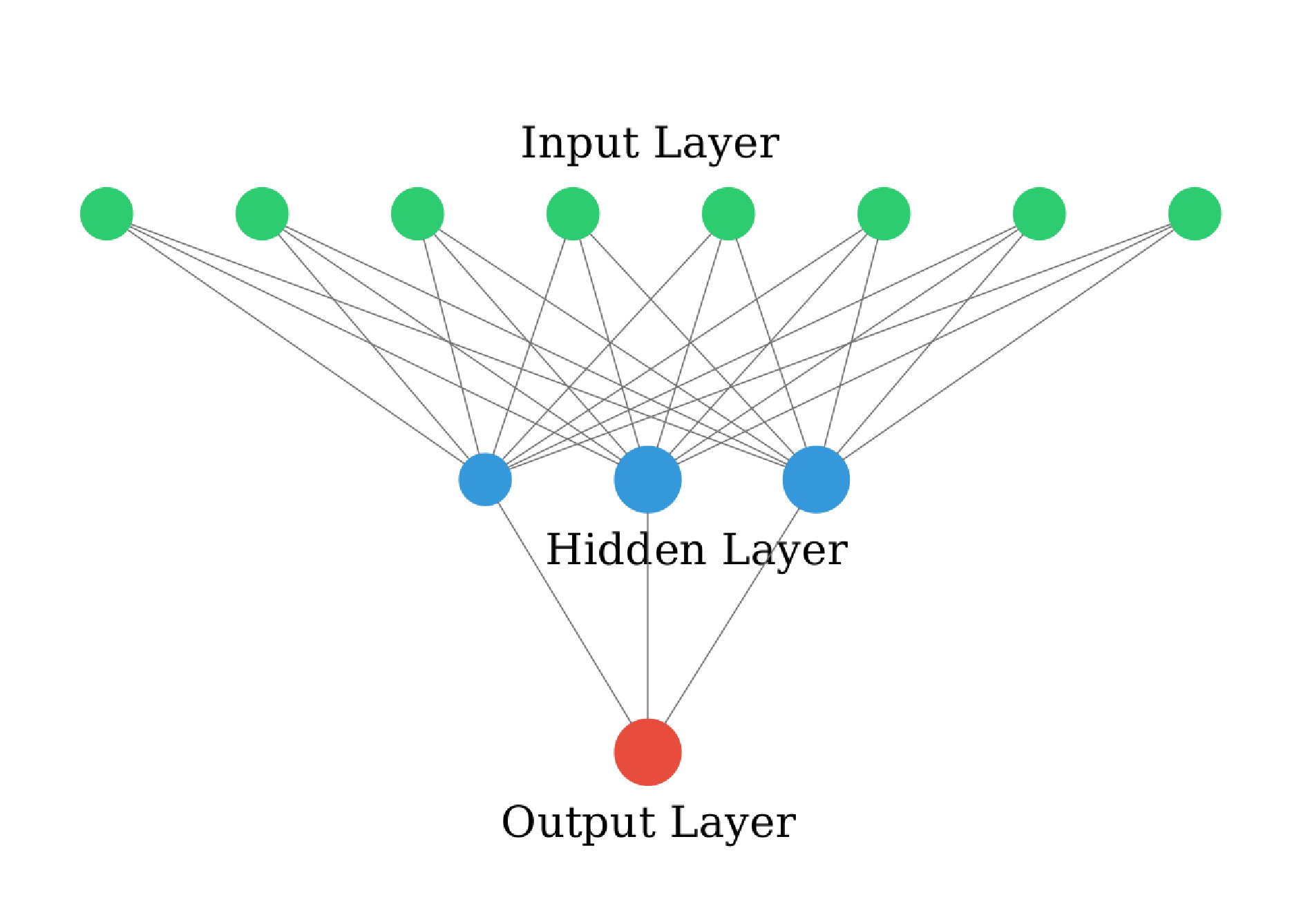}
  \caption{An example of a basic architecture of a neural network.}
\label{Fig: simple_ANN}
\end{figure}

Deep learning is part of a larger family of machine
learning methods based on artificial neural networks.
There are three types of learning for deep learning methods:
supervised, semi-supervised, and unsupervised.
Artificial neural networks ({\it ANNs}) were inspired by biological
systems' neuron networks. A typical {\it ANN} has
numerous layers between the input and output layers, as
shown in figure~(\ref{Fig: simple_ANN}), which depicts a neural network
with an input, a hidden layer, and an output layer.
Neural networks come in a variety of shapes and sizes, but they all include
the same basic components: neurons, weights, biases, and activation functions.
These components function in a similar way to the human brain and may be
trained in the same way as any other conventional {\it ML} method.
While the terms Deep learning and artificial
neural networks are often used interchangeably in some literature, there
is a significant difference between the two.  Neural networks use neurons to
convey data in the form of input values and output values through
connections, while deep learning is connected with feature transformation
and extraction, which aims to build a relationship between stimuli and the
corresponding neural responses.  For the sake of simplicity, here we are
going to use the term deep learning in a broader sense, which includes
simple applications of neural networks.  The reader is advised that this
may not be a generally accepted practice.

Three types of neural networks are among the most used in astronomy:

\begin{enumerate}

\item Multilayer Perceptron ($MLP$).
\item Long short-term memory ($LTSM$).
\item Convolutionary Neural Networks ($CNN$)
\end{enumerate}

A $MLP$ is a type of feedforward $ANN$, and is typically composed by
a network of multiple layers of perceptrons.  A single perceptron
is a function that can decide whether or not an input, represented by a
vector of numbers, belongs to some specific class \citep{Freund_99}.
A feedforward $ANN$ is a network in which nodes do not form a cycle of
connections, but only propagate forward.  There are at least three levels of
nodes in an $MLP$: an input layer, a hidden layer, and an output layer
(see also figure~(\ref{Fig: simple_ANN})).
Each node, except for the input nodes, is a neuron with a
nonlinear activation function. Backpropagation is a supervised learning
technique used by $MLP$ during training \citep{Rosenblatt1963PRINCIPLESON}.
$MLP$ were very popular in the 1980s for problems like speech and image
recognition, but other methods like $CNN$ have been more commonly used
for the latter applications more recently \citep{2019MNRAS.486.4158D}.
The latest applications of $MLP$ are in the field of time series forecasting
(see \citet{brownlee_2020}).

$LSTM$ is a deep learning architecture based on an artificial recurrent
neural network ($RNN$).  Unlike normal feedforward neural networks,
$LSTM$ has feedback connections, which allows them to handle not only
individual data points (such as photos), but also complete data streams
(such as speech or video). Applications of $LSTM$ in astronomy include
time series forecasting \citep{brownlee_2020} and time series
anomaly detection \citep{Pankaj_2015}.

$CNN$ are regularized versions of multilayer perceptrons. Multilayer
perceptrons are typically completely connected networks, meaning that each
neuron in one layer is linked to all neurons in the following layer.
These networks' "complete connectedness" makes them vulnerable to data
overfitting. Regularization, or preventing overfitting, can be accomplished
in a variety of methods, including punishing parameters during training
(such as weight loss) or reducing connectivity (skipped connections, dropout,
etc.).  To create patterns of increasing complexity, $CNN$s
employ a different method of regularization: they take advantage of the
hierarchical structure in data and use smaller and simpler patterns imprinted
in their filters. $CNN$s have shown to be effective on difficult computer vision
issues, reaching state-of-the-art results on tasks like picture
classification while also serving as a component in hybrid models for
completely new problems like object localization, image captioning, and
more \citep{brownlee_2020}. Because of their effectiveness, they
are the most commonly used $ANN$ architecture for 
image detection of Solar System small bodies
\citep{2019MNRAS.486.4158D, 2021AJ....161..218D}.

Finally, more complex architecture involving combinations of $LSTM$ and $CNN$
are also possible and have been discussed in \citet{brownlee_2020}.
Deep learning {\it Python} routines are available in the {\it Tensorflow}
\citep{tensorflow2015-whitepaper} and {\it Keras} packages \citep{Chollet_2018}.
Because of their applications to image processing, Convolutional
Neural Networks ($CNN$) can also very efficiently run on
{\it graphics processing unit (GPUs)} \citep{5452452}, and, more recently,
have found applications in {\it Tensor Processing Units (TPUs)}
\citep{2021arXiv210210423Y}.  Interested readers can find more information
on the subject in the above references.

\section{Applications to Solar System small bodies}
\label{sec: dl_appl}

\begin{figure}
  \includegraphics[width=3.5in]{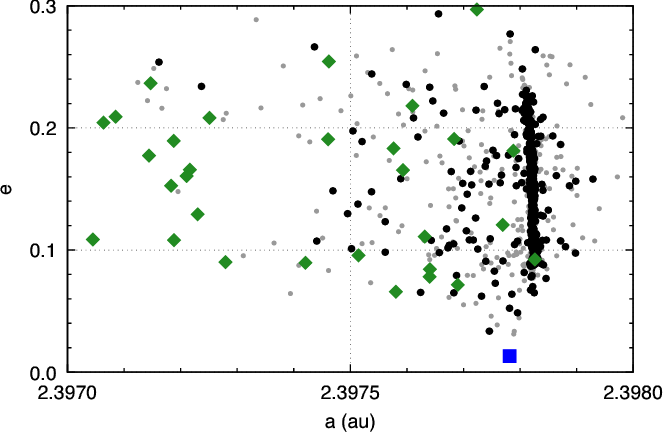}
  \caption{Asteroids in the resonances 4J-2S-1.  Grey dots
    are resonant asteroids from a previous catalog. Black
    dots are newly identified resonant asteroids according to
    the $ML$ method, blue squares are resonant asteroids not
    identified by $ML$, and green diamonds display false positive
    identification by $ML$. Adapted from figure~(1) by
    \citet{2017MNRAS.469.2024S} and reproduced with permission
    from the author and MNRAS (\copyright MNRAS).}
\label{Fig: smirnov}
\end{figure}

\citet{2017MNRAS.469.2024S} applied {\it KNN, Decision tree,
  Gradient boosting}, and {\it Logistic regression}  to identify
resonant three-body asteroids in the main belt.
The results of identification by machine learning methods were
accurate and take much less time than numerical integration (seconds
versus days).  The authors identified 404 new asteroids in the 4J-2S-1A
three-body resonance using a machine-learning methodology (see
figure~(\ref{Fig: smirnov})).

The subdivision of the observed objects of the  Kuiper Belt (KBOs) into
different dynamic classes is based on their orbits.
\citet{2020MNRAS.497.1391S} used a {\it GBoost} model to
automatically classify newly discovered objects into four
classes.

Some families of asteroids may also be the result of
spin-up-induced fission of a parent body in critical
rotation (fission clusters, \cite{2010Natur.466.1085P}).
In at least four young groups of fission, more than 5\% of its members
belong to subfamilies. {\it  K-means} can be used to automatically
identify these groups \citep{2020NatAs...4...83C}.

\citet{2019MNRAS.488.1377C} applied hierarchical
clustering algorithms {\it HCM} for supervised learning to identify
families of asteroids. This approach can find
family members with scores above 89.5\%. The authors identify 6 new
families and 13 new clusters in regions where the method can be applied.

\begin{figure}
  \includegraphics[width=3.5in]{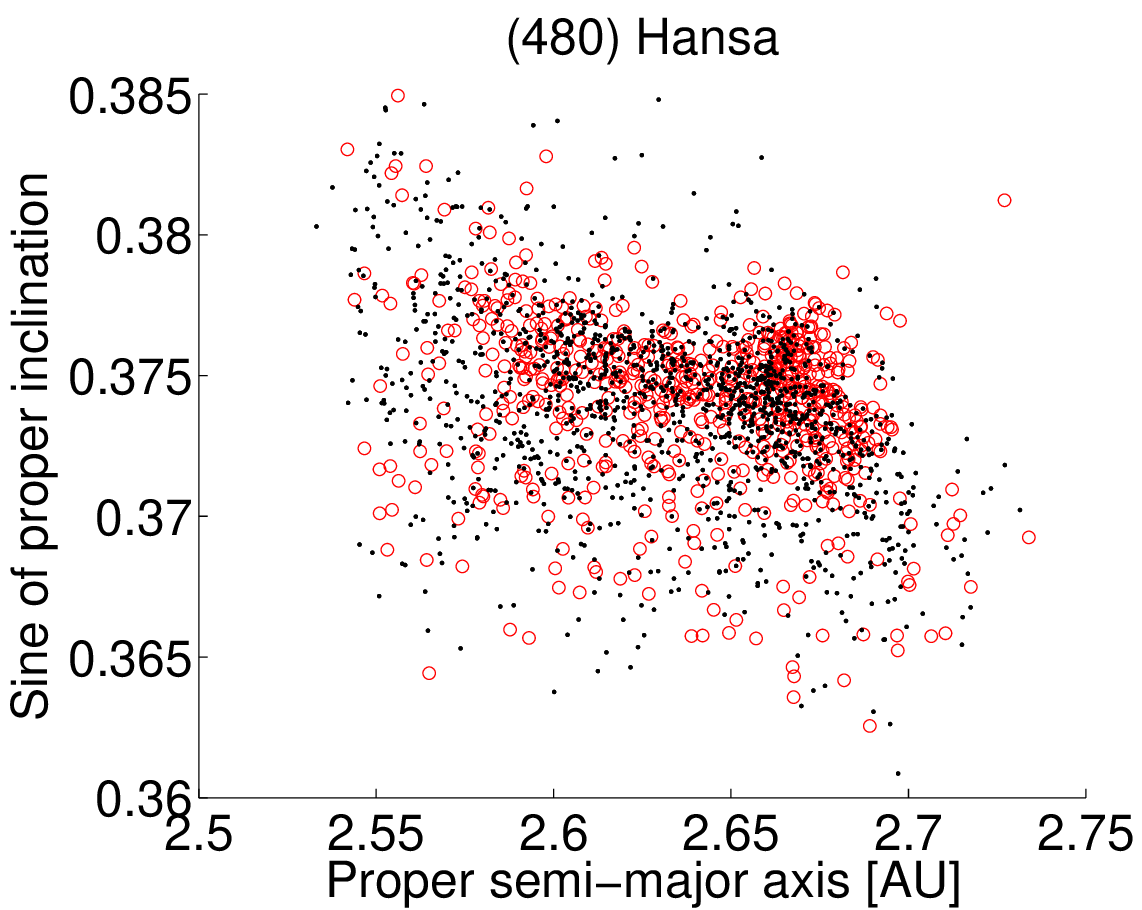}
  \caption{An $(a,\sin{(i)})$ projection of known members of
    the Hansa asteroid family. Asteroids predicted to be possible
    new members by {\it ExtraTree} are displayed as red circles.}
\label{Fig: Hansa}
\end{figure}

\citet{2020MNRAS.496..540C} used machine learning
classification algorithms to identify new family members based on the
orbital distribution in $(a, e, sin (i))$ of previously known family
members.  They compared the result of nine classification algorithms for
autonomous and joint approaches. The {\it Extremely Randomized Trees
(ExtraTree)} method had the highest precision, allowing to
recover up to 97\% of the family members identified with the
standard HCM method.  An application of this method to the Hansa asteroid
family is shown in figure~(\ref{Fig: Hansa}), where previously known
family members are shown as black dots, and members predicted by
the {\it ExtraTree} approach are red circles.

Selecting the ideal machine learning model and its hyperparameters can be a
difficult task, especially when the domain of hyperparameters to be
optimized is vast. Genetic algorithms \citep{Peng-Wei_2004} have this name
because they mimic the process of genetic evolution. They have
recently been used to select the ideal algorithm for predicting the asteroid
population in the secular resonances $z_1$ and $z_2$, which are resonances
involving the commensurabilities $g-g_6+s-s_6$ and $2(g-g_6)+s-s_6$, where
$g$ is the precession frequency of the longitude of pericenter, $s$ is
the precession frequency of the longitude of nodes, and the suffix 6 identify
Saturn, the sixth planet \citep{2021CeMDA.133...24C}.

If we extend the field to general applications to asteroids, including
observations, imaging, and taxonomy, three more works were produced
in the last few years.

\citet{2021RNAAS...5..199D} used a logistic Bayesian additive regression
tree model ($BART$) to find correlations between relative large
flux uncertainties and systematic right ascension errors
and large diameters in the {\it Gaia} database.

Taxonomic classification of NEA \citep{2017AJ....154..162E} and MBA
\citep{2018ApJS..237...19E} based on {\it VRI} spectrophotometry data was
recently obtained with algorithms using {\it KNN, SVM}, and a
{\it Gaussian Naive Bayes} method developed by \citet{2016AJ....151...98M}.

\citet{2018PASJ...70S..38C} used an algorithm developed by
\citet{2018PASJ...70S..39L} and based on {\it Random forest} and
{\it Isolation forest} for the supervised and unsupervised learning to detect
moving objects, including asteroids, from images obtained from the Hyper
Suprime-Cam Subaru Strategic Program (HSC-SSP).

Asteroid' brightness is not generally constant, but presents time
variations that depend on the asteroid shape, albedo, rotation
period, and other factors.  {\it ARIMA} models can fit the
light curve shapes with an error of less than 1\%, even for very complex
shapes. Based on the order of the best-fitting model $(p,d,q)$, a new
classification scheme for asteroid light curves can be introduced
\citep{Carruba_2021_l}.

\begin{figure}
   \includegraphics[width=3.0in]{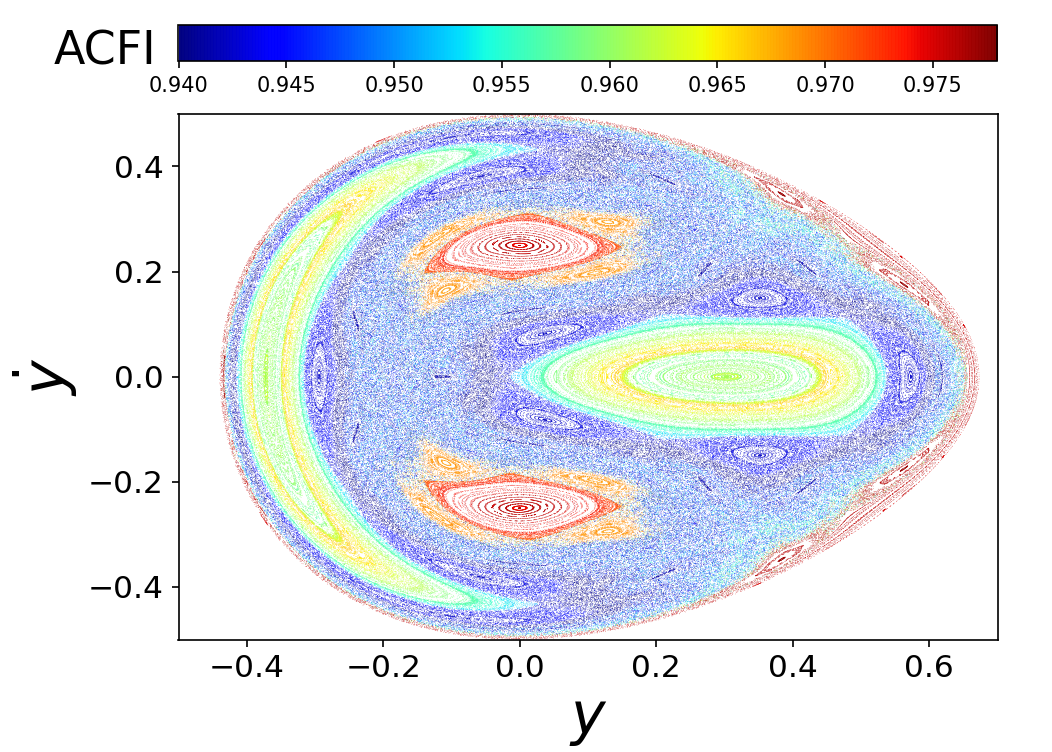}   
   \caption{A 2D Poincar\'{e} Surface of Section ($PSS$) with for the
     H\'{e}non-Heiles system with $H = 0.125$ and $ACFI$ values.
     Adapted from figure~(4) of \citep{2021CeMDA.133...24C}.}
\label{Fig: henon_heiles}
\end{figure}
     
$ACFI$ can be applied to well-known systems, like the H\'{e}non-Heiles
Hamiltonian (see figure~(\ref{Fig: henon_heiles})).
For Solar System small bodies, it can be used to
separate the effects of resonance overlapping from close encounters
with massive bodies \citep{2021CeMDA.133...24C}.

Neural networks can be used to predict the orbital stability of asteroids in the
main belt.  \citet{2021MNRAS.502.5362L} used {\it ANN} to fit data obtained
from the integration of 151,000  orbital initial conditions. A temporal
stability map in the $(a,e)$ plane was obtained as a result.

\begin{figure*}
  \includegraphics[width=6.5in]{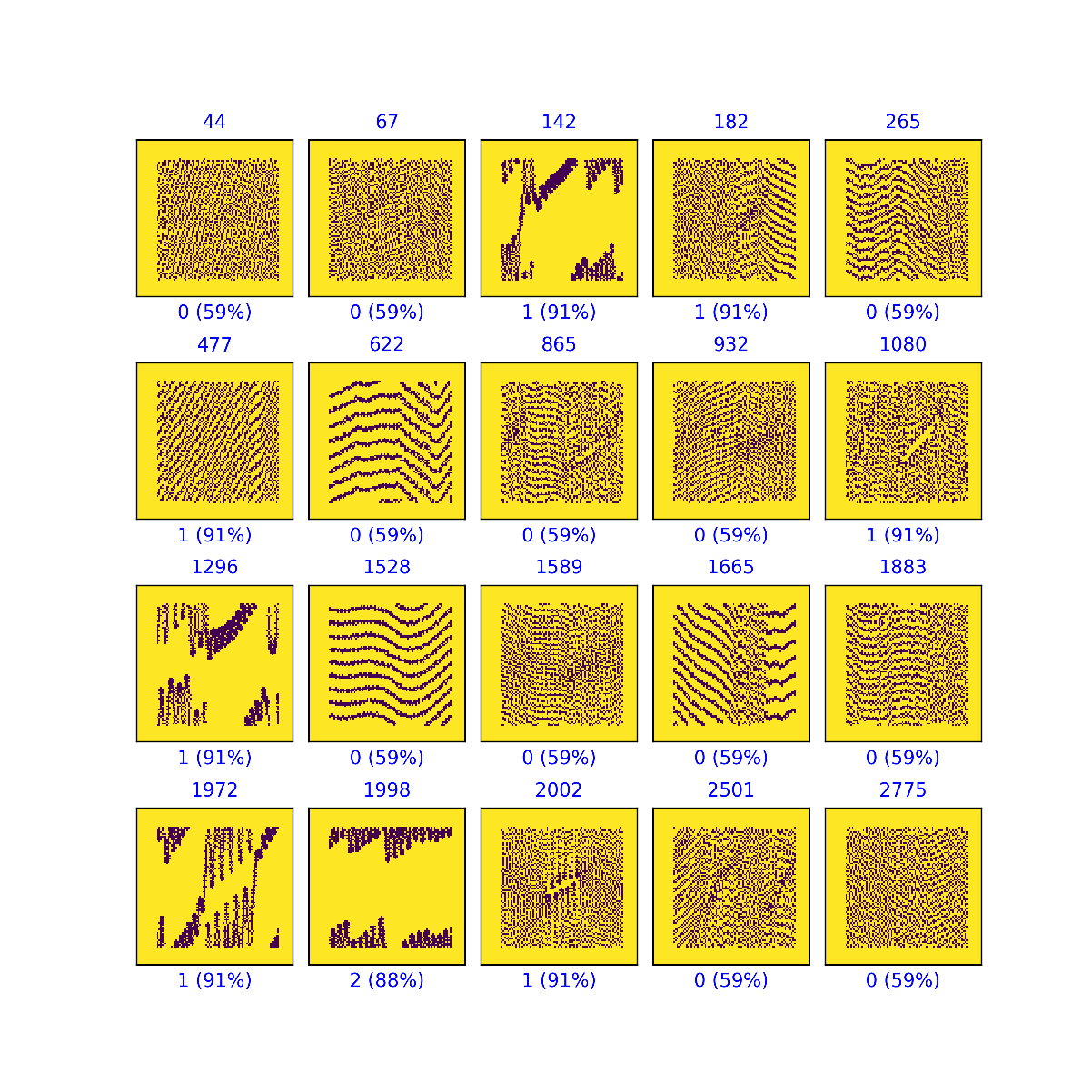}
  \caption{A set of 20 images of asteroids in or near resonant states
    of the 1:2 MMR (mean-motion resonance) with Mars. The x-axes display
    the predicted classification (0 for circulating orbits, 1 for
    switching and 2 for librating ones) and their probabilities of
    belonging to a given class.  The titles of each figure panel are
    the asteroid identifications.}
\label{Fig: Three_ANN}
\end{figure*}

Neural networks can also be used to automatically recognize images from
time series. This allows one to automatically determine
whether an asteroid is in resonance or not, as performed by
\citep{2021MNRAS.504..692C} for asteroids near the $M1:2$ mean-motion
resonance. Figure~(\ref{Fig: Three_ANN}) displays a set of 20 images of
resonant arguments for 20 such asteroids.  The first panel displays the
time behavior of an argument for an asteroid in a circulating orbit, the
seventeenth one that of an asteroid in a librating state, and the third
panel shows the case of an asteroid alternating phases of libration and
circulation. Similar sets of images can be used to train artificial neural
networks so that the process of identifying the resonant state is performed
automatically.

\citet{aljbaae_2021} applied a Time-Series prediction with Neural Networks
in {\it Python} with {\it Keras} to identify non-chaotic behavior of
spacecraft close to the asteroid (99942) Apophis during its exceptionally
close approach with Earth on 2029 April 13, at about 38 000 km from the
Earth’s center. The authors classify orbits based on a relationship between
the difficulty in the prediction and the stability. This method can
isolate the most regular orbits that could be stable in the system.
A good correlation was found between the Time-Series prediction approach and
{\it MEGNO} \citep{cincotta_2000} or the Perturbation Map of type II
(PMap) \citep{sanchez_2017, sanchez_2019}.
\citet{pugliatti_2020} adopted a CNN approach for the small-body
shape recognition task. The authors obtained very good performances of
their model, with a precision of 98.52\%. Meaningful results are also
presented with varying illumination conditions. 

Finally, \citet{2022MNRAS.511.2218L} recently used $ANN$ to
predict the trajectories of the 2:3 Neptune resonators over periods
of $\simeq$ 20000 yrs.  The outcome of the model can predict the
resonant angle with an accuracy as small as of a few degrees.
This approach, faster that standard numerical methods, can help
in identifying populations of KBOs to be discovered in future surveys.

If we extend the field to include the detection and imaging of small bodies,
a series of related papers have been published in the last years. 
\citet{2019MNRAS.486.4158D} created {\it DeepStreaks} in 2019, a
convolutional neural network, deep-learning system designed to efficiently
identify streaking fast-moving near-Earth objects identified in the
Zwicky Transient Facility (ZTF), a wide-field, time-domain survey.
The method cut human involvement in the streak detection process by
orders of magnitude, from several hours to less than 10 minutes every day. 

\begin{figure}
  \includegraphics[width=3.5in]{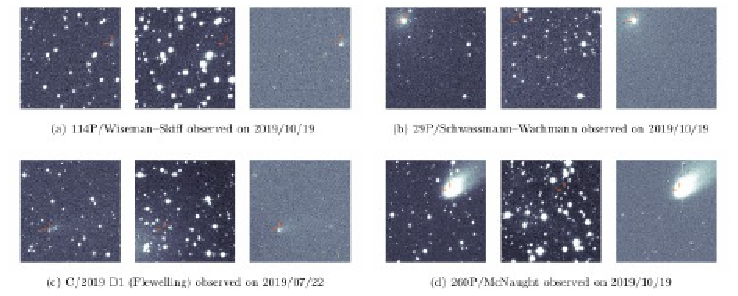}
  \caption{A set of images used by {\it Tails} for the purpose of training
    $ANN$s to recognize comets image data. Image adapted from the figure (1) of
    \citet{2021AJ....161..218D} and reproduced with permission of the American
    Astronomical Society (\copyright AAS).}
\label{Fig: Comet_ANN}
\end{figure}

\citet{2021AJ....161..218D} also developed {\it Tails},
an open-source deep-learning framework for the identification and
localization of comets in the image data of the Zwicky Transient Facility (ZTF).
{\it Tails} employs a custom EfficientDet-based architecture and is capable
of finding comets in single images in near-real-time
(see figure~(\ref{Fig: Comet_ANN}) for an example of images used to train
the {\it ANN}.  Information on the {\it ANN} architecture is available in figure
(1) of \citet{2021AJ....161..218D}).
This includes the first AI-assisted discovery of a comet (C/2020 T2) and
the recovery of a comet (P/2016 J3 = P/2021 A3).

Neural networks can also be used to automatically classify asteroids
into their taxonomic spectral classes. \citet{Pentilla_2021} used
feed-forward neural networks to predict asteroid taxonomies with the Bus-DeMeo
classification using datasets simulating {\it GAIA} observations,
and, more recently, simulating Legacy Survey of Space and Time (LSST)
observations \citep{10.3389/fspas.2022.816268}.
Results show that neural networks can identify taxonomic classes of
asteroids in a robust manner.

\section{From classification to insight: assessing the impact of
machine learning applications to asteroid dynamics}
\label{sec: class_insight}

\citet{2020WDMKD..10.1349F} utilized seven categories to assess recently
published literature, where the last two are those with a higher scientific
outcome.

\begin{enumerate}

\item {\bf Classification}: objects or features are assigned to one of
  several categories or labels.  The machine learning algorithm learns the
  properties that link an instance to a category based on a training set
  (labeled or unlabeled).  When the algorithm is applied to a new instance,
  it assigns the most likely category label. 

\item {\bf Regression}: The assignment of a numerical value (or values)
  based on the characteristics that the machine learning algorithm has
  learned or otherwise predicted. A training set, similar to classification,
  can be utilized, or the features might be deduced from the data set.

\item {\bf Clustering}: these techniques assess if an object or feature
  is a part of (or a member of) anything. This could be a physical structure
  or association, like an asteroid family, or a region within an
  N-dimensional parameter space. 

\item {\bf Forecasting}: The machine learning algorithm's goal is to
  learn from prior occurrences and anticipate or forecast the occurrence
  of a similar event. The forecast has an inherent time dependence. 

\item {\bf Generation and Reconstruction}:  Missing information is generated
  and reconstructed, with the expectation that it will be consistent with
  the underlying truth.  The absence of information could be caused by noise,
  processing artifacts, or other astronomical events, all of which conspire
  to obfuscate the required signal. 

\item {\bf Discovery}: New celestial objects, features, or relationships
  are discovered as a result of using a machine learning or artificial
  intelligence technology. 

\item {\bf Insight}: New scientific knowledge is demonstrated as a result
  of applying machine learning, which goes beyond the finding of
  astronomical objects. This includes situations where information about
  the suitability of using machine learning, data set selection,
  hyperparameters, and comparisons to human-based classification is gathered. 
  
\end{enumerate}

Classification, regression, and clustering methods are frequently compared
to a comparable human-centered approach, but with the need to “scale-up”
either the size of the dataset to be studied or the time it takes to
complete the work.  The results of classification and regression can be
the culmination of an investigation or the input to a forecasting,
generating, finding, or insight process. 

These seven categories, which constitute a loose hierarchy of
sophistication, provide an assessment of the maturity of the use of
machine learning and artificial intelligence within a subfield of astronomy.
Using a machine learning technique to conduct a classification, regression,
or clustering problem is a frequent starting point.
Machine learning can be used to estimate expected future outcomes,
like future solar flares \citep{2018SoPh..293...28F}, or produce 
discoveries, like the identification of new candidates of rare objects
\citep{2018AJ....155..108Z}, once it has been established as being
comparable to, or exceeding, a more traditional approach. 

How can papers published in the field of asteroid dynamics and
Solar System small bodies be classified
in terms of these categories?  We can divide the field into five
main types of subfields:

\begin{enumerate}

\item{\bf Resonant and chaotic dynamics}:  in this subfield we consider papers
  that deal with the identification and classification of asteroidal populations
  in mean-motion and secular resonances, or deal with the effects of
  chaotic motions either caused by resonant dynamics or stochastic effects,
  like close encounters with massive bodies.

\item{\bf Asteroid families' identification}: articles that deal with
  identifying new members of known or newly identified asteroid groups,
  which could either be dynamical families or groups produced by rotational
  fission.

\item{\bf $ML$ applications to time series analysis}: articles that focus on
  the prediction and classification of a time series, which could be a time
  series of orbital elements, an asteroid light curve, or other applications.
   
\item{\bf Image recognition}: these are papers that mostly use {\it Artificial
  Neural Networks (ANN)} to recognize patterns or images.  Applications include
  asteroid or comets images, asteroid spectra, and asteroid resonant arguments,
  among others.
  
\item{\bf Asteroid taxonomy and physical properties}: papers that focus on
  obtaining taxonomical information from spectral or spectrophotometric data,
  or physical properties like masses, shapes, and densities, using $ML$.

\end{enumerate}

These categories are not mutually exclusive.  For instance, papers that
deal with resonant dynamics can use {\it ANN} to recognize images of
resonant arguments \citep{2021MNRAS.504..692C}. Articles that focus
on chaos may use time series analysis techniques to identify regular and
chaotic orbits \citep{2021CeMDA.133...24C}.  Once the number of papers
in the field increases, other divisions into subfields may appear to be
more useful for classifying research done in this area. For the time being,
however, we will use this scheme to organize current papers in the field.

\begin{figure}
  \includegraphics[width=3.5in]{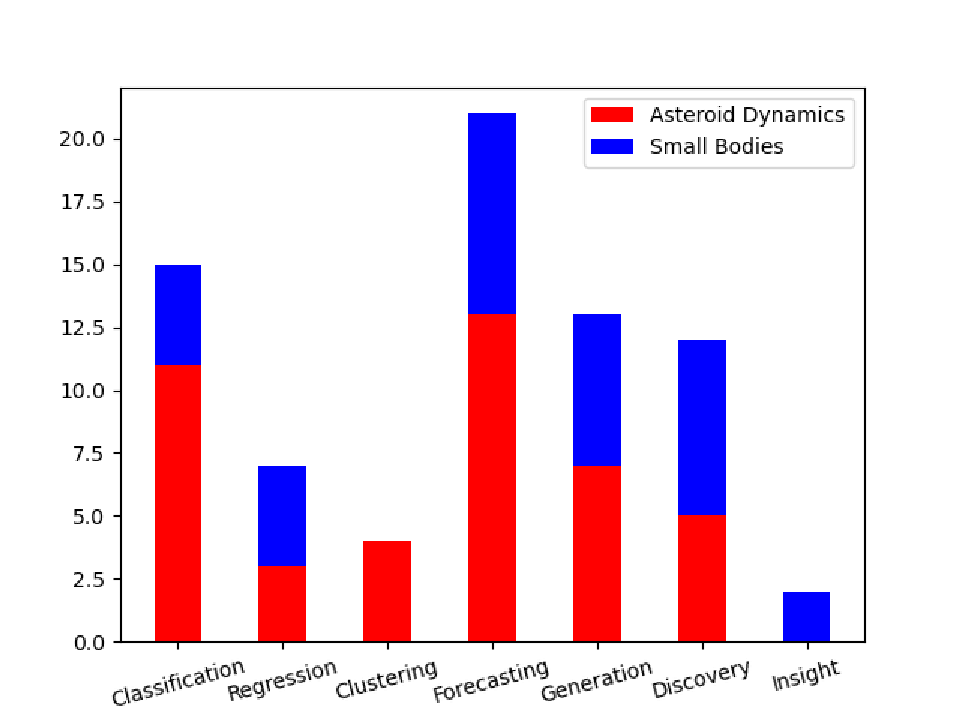}
  \caption{A bar plot summarizing the data from table~(\ref{Table: qual_ast})
    for the categories of $ML$ algorithms applied to asteroid dynamics
    (blue bars) and small bodies (red bars).}
\label{Fig: Bar_Fluke}
\end{figure}

Table~(\ref{Table: qual_ast}) displays our analysis of 12 papers in the
field of machine learning applied to asteroid dynamics, and 21 in the broader
area of Solar System small bodies. For each of the sub-areas that we
introduced in this section, we classify if a paper covers one or more of the
seven categories of scientific outcomes introduced by
\citep{2020WDMKD..10.1349F}.  Results are also summarized for the
whole subareas of asteroid dynamics and small bodies in
figure~(\ref{Fig: Bar_Fluke}).  A more detailed classification, article by
article, is also presented in table~(\ref{Table: ref_class}) in 
appendix (1).

\begin{table*}
  \begin{center}
    \caption{A qualitative summary of the categories of $ML$ algorithms
      and the most common subareas of research for asteroid dynamics (first
      four rows) and for the extended area of applications to small bodies
      (second set of five rows).}
        \label{Table: qual_ast}
         \begin{tabular}{|c|c|c|c|c|c|c|c|}
           \hline
             
{\bf Nature/type} & {\bf Classification} & {\bf Regression} & {\bf Clustering} & {\bf Forecasting} & {\bf Generation} & {\bf Discovery} & {\bf Insight} \\
\hline
 {\bf Res. \& Chaotic Dyn} & 6 & 2 & 1 & 7 & 3 & 2 & 0 \\
    {\bf Ast. Fam. Id.}    & 1 & 0 & 2 & 3 & 1 & 2 & 0 \\
    {\bf Time series An.}  & 3 & 1 & 1 & 3 & 2 & 1 & 0 \\
    {\bf Image recognition}& 1 & 0 & 0 & 1 & 1 & 0 & 0 \\
\hline
 {\bf Res. \& Chaotic Dyn} & 6 & 3 & 1 & 7 & 3 & 3 & 0 \\
    {\bf Ast. Fam. Id.}    & 1 & 0 & 2 & 3 & 1 & 2 & 0 \\
    {\bf Time series An.}  & 3 & 2 & 1 & 4 & 4 & 2 & 0 \\
    {\bf Image recognition}& 1 & 1 & 0 & 4 & 4 & 4 & 2 \\
    {\bf Ast. Tax. Phys}   & 4 & 1 & 0 & 4 & 1 & 1 & 0 \\
    \hline
\end{tabular}
\end{center}
\end{table*}

What can be learned from this analysis? \citet{2020WDMKD..10.1349F} used a
hierarchy of three categories to evaluate the maturity of $ML$ within a
subfield of astronomy: {\it emerging, progressing}, and {\it established}.

The {\it emerging} stage is used for astronomy and astrophysics subfields
that are beginning to study the use of machine learning and artificial
intelligence, generally by attacking easier to solve problems.
This could be a challenge that necessitates a classification or
regression strategy, or a comparison of machine learning and another,
well-known method. For {\it emerging} communities, the 
$ML$ methodologies best suited to a given problem may have not yet
been encountered, or the community is tiny. 

In a {\it progressing} stage, a greater range of approaches are utilized,
or a single technique is used several times, or there is an immediate jump
to the forecasting, discovery, or insight phases in fields regarded as
progressing in their usage of ML.

Finally, a subfield reached an {\it established} stage if the application
of machine learning  has become essential in
established subfields, there is a considerable body of research, and the
focus is mostly on predicting, discoveries, or insight.
It is no longer necessary to assess the applicability of machine learning
because its application has become of wide use.

\citet{2020WDMKD..10.1349F} classified the study of {\it Small bodies}
objects as having reached a {\it progressing} stage. Based on our
analysis, we agree on this result: there is a wider community already
studying {\it Small bodies}, which has produced works that
produced scientific outcomes reaching the categories of {\it discovery}
and {\it insight}, like the articles from \citet{2019MNRAS.486.4158D} and
\citet{2021AJ....161..218D}, with the first $ML$-assisted discovery of
a comet (C/2020 T2).  More advanced techniques, like the use of neural
networks for image recognition, are becoming more and
more common in the area, almost reaching the status of
an established method, associated with the \citep{2020WDMKD..10.1349F}
{\it progressing} stage.

This is less true if we consider the subfield of {\it asteroid dynamics}.
The community working in the area is less established, with six
groups with published papers so far, 
fewer papers produced scientific outcomes that can be categorized as belonging
to {\it generation} or {\it discovery}, and none reached the stage of
{\it insight}.  Overall, we believe that $ML$ applied to asteroid dynamics
may be classified as still being in the {\it emerging} phase, but already
approaching a {\it progressing} phase.

\section{Future trends}
\label{sec: future trends}

Having reviewed recent literature in the field, it is now time to
attempt to look for possible future trends in the area.  In performing
such a task one always runs the risk of painting a biased portrait, influenced
by personal experiences and expectations.  Yet, we think that this could
be a useful thought experiment, and could at least form the base for
a more educated discussion on the subject.

Most recent efforts have focused on the use of {\it ANNs},
either for classification of images of asteroid resonant arguments
\citep{2021MNRAS.504..692C}, for the detection of images of Solar System
small bodies \citep{2019MNRAS.486.4158D, 2021AJ....161..218D}, or for
classifying asteroids into their taxonomic spectral classes
\citet{Pentilla_2021, 10.3389/fspas.2022.816268}.  The use
of more and more complex architecture of {\it ANNs}, as described
in section~(\ref{sec: Deep learning}), could significantly
improve the speed and efficiency of the models currently used
by the scientific community.  It is reasonable to expect
that some of the next scientific advancements may occur
in this area.

An engine for progress in the field of $ML$ has been in the recent past
Kaggle competitions (see also https://www.\\kaggle.com/competitions,
last accessed on June 1st 2022).
Kaggle competitions are machine learning tasks made
by Kaggle or other companies, like Google or WHO. Competitions range in
types of problems and complexity, but they are generally open to novices.
They usually award a monetary prize for the best problem solution.
A recent example of a Kaggle-like competition in the field of asteroid
dynamics has been a challenge hosted in Kelvins about the prediction
of the effects of an impact while deflecting asteroids in the context of
the HERA mission \\ (https://kelvins.esa.int/\-planetary-defence/, last accessed
on August 5th, 2021). A dataset is
provided to explore a way to do this task in a purely $ML$-driven way.

Another way in which the field could advance is by using algorithms that
have found applications in other fields, but are not yet of common use
among the small bodies community.  Examples of these are density-based
clustering algorithms, like {\it DBSCAN} and {\it GDBSCAN}, which can
cluster point objects, as well as spatially extended objects according to
both, their spatial and their non-spatial attributes \citep{DBSCAN_1998}. 

Concerning deep learning algorithms, a very promising method is the
{\it Transformer architecture} \citep{vaswani2017attention}.  
Complex recurrent or convolutional neural networks with an encoder and a
decoder are the most common sequence transduction models. The finest models
additionally use an attention mechanism to connect the encoder and decoder.
The Transformer is a new simple network architecture model purely based
on attention processes, with no recurrence or convolutions.
Experiments on two machine translation tasks reveal that these models are
superior in quality, parallelizable, and take much less time to train.

Another area where applications of $ML$ are most likely to be of
great importance is the analysis of data produced by next-generation
astronomical surveys. We already saw that large databases produced by
space missions like {\it GAIA} will need $ML$ methods to fully take advantage
of the information available in the data \citep{Pentilla_2021}.
Ongoing observational robotic surveys, like those produced by the
Zwicky Transient Facility ($ZTF$) \citep{2020PASP..132c8001D} have already
produced modeling efforts in the $ML$ area, with the already discussed
papers on the {\it DeepStreaks} \citep{2019MNRAS.486.4158D} and {\it Tails}
\citep{2021AJ....161..218D} software packages for the identification of
fast-moving near-Earth objects and comets.  $ZTF$ is a
large optical survey in multiple ﬁlters producing hundreds of thousands of
transient alerts per night.  $ML$ is already commonly used for the
analysis of such data, and plans to expand its implementations are already
underway \citep{Mahabal_2019}.

Data produced by $ZTF$ is, however, estimated to be just $\simeq$ 10\%
of what is expected from the Vera C. Rubin Observatory.  This 8.4-meter
three-mirror design telescope, currently under construction in Chile, is
expected to start operations in October 2023.  \citet{Lynne_2015} estimated
that the catalogs generated by the Vera C. Rubin Observatory will increase
the known number of small
bodies in the Solar System by a factor of 10-100 times, among all
populations, with $\simeq$ 1 million new asteroid discoveries predicted
in just the first year of operations.  The need to handle such a massive
amount of data has already caused the development of new methods, like
the use of Lomb-Scargle algorithm on Graphics Processing Units
\citep{Gowanlock2021FastPS}.  More $ML$-oriented software development
is expected in the next years.

\section{Conclusions}
\label{sec: concl}

In this work, we reviewed 21 papers with applications of $ML$ to Solar System
bodies, 12 of which had a focus on $ML$ applied to asteroid dynamics.
A brief outline of the methods used in the reviewed references has been
first provided, with references to the theoretical papers most relevant
to each approach, and links to some of the most commonly used software
packages for each method.  Current references were then classified in
terms of metrics recently used by \citet{2020WDMKD..10.1349F} to assess
the impact of applications of $ML$ algorithms in other
astronomical subfields.

$ML$ applications to Solar System bodies, which include imaging and
spectophotometry of small bodies have already reached the {\it progressing}
stage, with established research communities and
methodologies, as well as papers describing how the use of $ML$ led
to the discovery of new celestial objects, or features, or to new insights
in the field. $ML$ applied to asteroid dynamics has seen many recent
developments, especially in the Space missions to asteroids area.
However, since fewer papers producing discoveries or insights are still
produced, we believe that this area may be still considered as an {\it
  emerging} field, moving on to the {\it progressing} phase.

As discussed in section~(\ref{sec: future trends}), the most recent
efforts have focus on the use of {\it ANNs},
either for classification of images of asteroid resonant arguments,
for the detection of images of Solar System small bodies, or for
classifying asteroids into their taxonomic spectral classes
Developments on the line of what was discussed in
section~(\ref{sec: Deep learning}), could in the next year
significantly improve the speed and efficiency of
methods currently used by the scientific community.

Finally, the development of new, more advanced, $ML$ algorithms, and the
need to develop new methods able to handle the massive amount of data expected
to be produced by robotic astronomical surveys, like the
Zwicky Transient Facility ($ZTF$) and the Vera C. Rubin Observatory have
already been a powerful motivation to develop new $ML$ methods for data
analysis, as discussed in section~(\ref{sec: future trends}).  It is fair
to expect many new exciting developments and discoveries in the field
in the next few years.

\section{Appendix (1): data on current applications of $ML$ in asteroid dynamics}
\label{sec: app_1}

\begin{table*}
  \begin{center}
    \caption{Reference data used to classify the status of application of
      machine learning to recent literature. The first column
      shows the article number, which refers to the order of citation in
      this work.  The second column
      yields the reference to the cited work. The third
      column displays the types of machine learning applications, which could
      be either an application of supervised (SL) or unsupervised (UL) learning,
      time-series analysis (TSA), or deep learning (DL). The fourth column
      show if the article is in the field of asteroid dynamics (AD) or
      small bodies in the Solar System (SB).
      The fifth column shows the article field, which can be resonant and
      chaotic dynamics (RD), asteroid families'identification (AFI),
      time-series analysis (TSA), Image recognition (IR), asteroid
      taxonomy and physical properties (ATP), or space missions to
      asteroids (ASM). Finally, the sixth column shows the
      \citet{2020WDMKD..10.1349F} categories by which
      the scientific outcomes of the paper were classified in this work:
      Classification (C), Regression (R), Clustering (CL), Forecasting (F),
      Generation and Reconstruction (G), Discovery (D), and Insight (I).
      Multiple categories may apply to a single work.}
    \label{Table: ref_class}
    \begin{tabular}{|c|c|c|c|c|c|}
      \hline
Article &  Reference   &   ML & Article & Subject   & Category \\
number  &    &   application  & Field   & Subfield &          \\
      \hline
  1 & \citet{2017MNRAS.469.2024S} & SL & AD & RD & C,R,F \\
  2 & \citet{2020MNRAS.497.1391S} & SL & AD & AFI& R,CLF \\
  3 & \citet{2020NatAs...4...83C} & UL & AD & AFI& CL,F \\
  4 & \citet{2019MNRAS.488.1377C} & UL & AD & AFI& CL,F \\
  5 & \citet{2020MNRAS.496..540C} & SL & AD & AFI& C,F,D \\
  6 & \citet{2021CeMDA.133...24C} & SL & AD & RD,AFI & R,F,G,D\\
  7 & \citet{2021RNAAS...5..199D} & SL & SL & ATP & C,F \\
  8 & \citet{2017AJ....154..162E} & SL & SL & ATP & C,F \\
  9 & \citet{2018ApJS..237...19E} & SL & SL & ATP & C,F \\
 10 & \citet{2018PASJ...70S..38C} & UL & SL & IR & R,D,I\\
 11 & \citet{Carruba_2021_l}      & TSA& AD & TSA& C,R,CL,F,G\\
 12 & \citet{2021CeMDA.133...24C} & TSA& AD & RD,TSA& C,F,D \\
 13 & \citet{2021MNRAS.502.5362L} & DL & AD & RD & C,F,G \\
 14 & \citet{2021MNRAS.504..692C} & DL & AD & RD,IR & C,F,G\\
 15 & \citet{aljbaae_2021}        & DL & AD & RD,TSA & C,F,G \\
 16 & \citet{2019MNRAS.486.4158D} & DL & SL & IR & F,D,G \\
 17 & \citet{2021AJ....161..218D} & DL & SL & IR & F,D,G,I\\
 18 & \citet{Pentilla_2021}       & DL & SL & ATP & C,F,G \\
 19 & \citet{10.3389/fspas.2022.816268} & DL & SL & ATP & C,F,G \\
 20 & \citet{pugliatti_2020}      & DL & AD & ATP & C,G \\
 21 & \citet{2022MNRAS.511.2218L} & DL & SL & RD, TSA & R,F,G,D \\
 \hline
\end{tabular}
\end{center}
\end{table*}

In this section we report the data used to classify current literature
according to \citet{2020WDMKD..10.1349F} category scheme, as discussed
in section (\ref{sec: class_insight}).  Our data is shown in
table~(\ref{Table: ref_class}).

\section*{Acknowledgments}

We are grateful to two anonymous reviewers for helpful and
constructive comments that much increased the quality of this work.
We would like to thank the Brazilian National Research Council
(CNPq, grant 304168/2021-1), the São Paulo Research Foundation (FAPESP,
grant 2016/024561-0), and the Institutional Training program (PCI/INPE,
subproject 6.8.1, Public Call 01/2021).
We are grateful to Dr. E. Smirnov and Dr. D. Duev for allowing us to use
figures from their papers \citep{2017MNRAS.469.2024S, 2021AJ....161..218D},
and to Dr. E. Smirnov for reading a preliminary version of this paper
and for useful comments and suggestions.
VC and RC are part of "Grupo de Din\^{a}\-mica Orbital \& Planetologia (GDOP)"
(Research Group in Orbital Dynamics and Planetology) at UNESP, campus
of Guaratin\-guet\'{a}. This is a publication from the MASB
(Machine-learning applied to small bodies, \\
https://valeriocarruba.github.\-io/Site-MASB/) research group.  Questions on
this paper can also be sent to the group email address:
{\it mlasb2021@gmail.com}.

\section*{Conflict of interest}
The authors declare that they have no conflict of interest.

\section{Author contributions}

All authors contributed to the study conception and design. Material
preparation, and data collection were performed by Valerio Carruba,
Safwan Aljbaae and Rita Cassia Domingos. The first draft of the manuscript
was written by Valerio Carruba and all authors commented on previous versions
of the manuscript. All authors read and approved the final manuscript.

\bibliographystyle{spbasic}      
\bibliography{mybib}

\begin{thebibliography}{77}
\providecommand{\natexlab}[1]{#1}
\providecommand{\url}[1]{{#1}}
\providecommand{\urlprefix}{URL }
\expandafter\ifx\csname urlstyle\endcsname\relax
  \providecommand{\doi}[1]{DOI~\discretionary{}{}{}#1}\else
  \providecommand{\doi}{DOI~\discretionary{}{}{}\begingroup
  \urlstyle{rm}\Url}\fi
\providecommand{\eprint}[2][]{\url{#2}}

\bibitem[{Abadi et~al.(2015)Abadi, Agarwal, Barham, Brevdo, Chen, Citro,
  Corrado, Davis, Dean, Devin, Ghemawat, Goodfellow, Harp, Irving, Isard, Jia,
  Jozefowicz, Kaiser, Kudlur, Levenberg, Man\'{e}, Monga, Moore, Murray, Olah,
  Schuster, Shlens, Steiner, Sutskever, Talwar, Tucker, Vanhoucke, Vasudevan,
  Vi\'{e}gas, Vinyals, Warden, Wattenberg, Wicke, Yu, and
  Zheng}]{tensorflow2015-whitepaper}
Abadi M, Agarwal A, Barham P, Brevdo E, Chen Z, Citro C, Corrado GS, Davis A,
  Dean J, Devin M, Ghemawat S, Goodfellow I, Harp A, Irving G, Isard M, Jia Y,
  Jozefowicz R, Kaiser L, Kudlur M, Levenberg J, Man\'{e} D, Monga R, Moore S,
  Murray D, Olah C, Schuster M, Shlens J, Steiner B, Sutskever I, Talwar K,
  Tucker P, Vanhoucke V, Vasudevan V, Vi\'{e}gas F, Vinyals O, Warden P,
  Wattenberg M, Wicke M, Yu Y, Zheng X (2015) {TensorFlow}: Large-scale machine
  learning on heterogeneous systems.
  \urlprefix\url{https://www.tensorflow.org/}, software available from
  tensorflow.org

\bibitem[{{Akhter} et~al.(2020){Akhter}, {Hassan}, and
  {Abbas}}]{2020A&C....3200403A}
{Akhter} MF, {Hassan} D, {Abbas} S (2020) {Predictive ARIMA Model for coronal
  index solar cyclic data}. Astronomy and Computing 32:100403,
  \doi{10.1016/j.ascom.2020.100403}

\bibitem[{{Aljbaae} et~al.(2021){Aljbaae}, {Souchay}, {Carruba}, {Sanchez}, and
  {Prado}}]{aljbaae_2021}
{Aljbaae} S, {Souchay} J, {Carruba} V, {Sanchez} DM, {Prado} AFBA (2021)
  {Influence of Apophis' spin axis variations on a spacecraft during the 2029
  close approach with Earth}. Accepted by the Romanian Astronomical Journal
  arXiv:2105.14001, \eprint{2105.14001}

\bibitem[{{Ball} and {Brunner}(2010)}]{2010IJMPD..19.1049B}
{Ball} NM, {Brunner} RJ (2010) {Data Mining and Machine Learning in Astronomy}.
  International Journal of Modern Physics D 19(7):1049--1106,
  \doi{10.1142/S0218271810017160}, \eprint{0906.2173}

\bibitem[{{Baron}(2019)}]{2019arXiv190407248B}
{Baron} D (2019) {Machine Learning in Astronomy: a practical overview}. arXiv
  e-prints arXiv:1904.07248, \eprint{1904.07248}

\bibitem[{{Bendjoya} and {Zappal{\`a}}(2002)}]{2002aste.book..613B}
{Bendjoya} P, {Zappal{\`a}} V (2002) {Asteroid Family Identification}, Arizona
  Univ. Press, pp 613--618

\bibitem[{Bezanson et~al.(2017)Bezanson, Edelman, Karpinski, and
  Shah}]{bezanson2017julia}
Bezanson J, Edelman A, Karpinski S, Shah VB (2017) Julia: A fresh approach to
  numerical computing. SIAM review 59(1):65--98,
  \urlprefix\url{https://doi.org/10.1137/141000671}

\bibitem[{Boehmke and Greenwell(2019)}]{Boehmke19}
Boehmke B, Greenwell B (2019) {Gradient Boosting". Hands-On Machine Learning
  with R}. Chapman \& Hall, London

\bibitem[{{Box} and {Jenkins}(1976)}]{1976tsaf.conf.....B}
{Box}, {Jenkins} (eds) (1976) {Time series analysis. Forecasting and control},
  San Francisco : Holden-Day

\bibitem[{Brownlee(2020)}]{brownlee_2020}
Brownlee J (2020) Deep Learning for Time Series Forecasting. Ed. Machine
  Learning Mastery, San Juan, PR, USA

\bibitem[{{Carruba} and {Aljbaae}(2021)}]{Carruba_2021_l}
{Carruba} V, {Aljbaae} S (2021) {Predicting asteroid lightcurves using ARIMA
  models}. In: European Planetary Science Congress, pp EPSC2021--36,
  \doi{10.5194/epsc2021-36}

\bibitem[{{Carruba} et~al.(2019){Carruba}, {Aljbaae}, and
  {Lucchini}}]{2019MNRAS.488.1377C}
{Carruba} V, {Aljbaae} S, {Lucchini} A (2019) {Machine-learning identification
  of asteroid groups}. \mnras 488(1):1377--1386, \doi{10.1093/mnras/stz1795}

\bibitem[{{Carruba} et~al.(2020{\natexlab{a}}){Carruba}, {Aljbaae}, {Domingos},
  {Lucchini}, and {Furlaneto}}]{2020MNRAS.496..540C}
{Carruba} V, {Aljbaae} S, {Domingos} RC, {Lucchini} A, {Furlaneto} P
  (2020{\natexlab{a}}) {Machine learning classification of new asteroid
  families members}. \mnras 496(1):540--549, \doi{10.1093/mnras/staa1463}

\bibitem[{{Carruba} et~al.(2020{\natexlab{b}}){Carruba}, {Spoto}, {Barletta},
  {Aljbaae}, {Fazenda}, and {Martins}}]{2020NatAs...4...83C}
{Carruba} V, {Spoto} F, {Barletta} W, {Aljbaae} S, {Fazenda} {\'A}L, {Martins}
  B (2020{\natexlab{b}}) {The population of rotational fission clusters inside
  asteroid collisional families}. Nature Astronomy 4:83--88,
  \doi{10.1038/s41550-019-0887-8}

\bibitem[{{Carruba} et~al.(2021{\natexlab{a}}){Carruba}, {Aljbaae}, and
  {Domingos}}]{2021CeMDA.133...24C}
{Carruba} V, {Aljbaae} S, {Domingos} RC (2021{\natexlab{a}}) {Identification of
  asteroid groups in the z$_{1}$ and z$_{2}$ nonlinear secular resonances
  through genetic algorithms}. Celestial Mechanics and Dynamical Astronomy
  133(6):24, \doi{10.1007/s10569-021-10021-z}

\bibitem[{{Carruba} et~al.(2021{\natexlab{b}}){Carruba}, {Aljbaae}, {Domingos},
  and {Barletta}}]{2021MNRAS.504..692C}
{Carruba} V, {Aljbaae} S, {Domingos} RC, {Barletta} W (2021{\natexlab{b}})
  {Artificial neural network classification of asteroids in the M1:2
  mean-motion resonance with Mars}. \mnras 504(1):692--700,
  \doi{10.1093/mnras/stab914}, \eprint{2103.15586}

\bibitem[{Chen et~al.(2004)Chen, Wang, and Lee}]{Peng-Wei_2004}
Chen PW, Wang JY, Lee H (2004) Model selection of svms using ga approach. 2004
  IEEE International Joint Conference on Neural Networks (IEEE Cat No04CH37541)
  3:2035--2040 vol.3

\bibitem[{Chen and Guestrin(2016)}]{Chen_2016}
Chen T, Guestrin C (2016) Xgboost. Proceedings of the 22nd ACM SIGKDD
  International Conference on Knowledge Discovery and Data Mining
  \doi{10.1145/2939672.2939785},
  \urlprefix\url{http://dx.doi.org/10.1145/2939672.2939785}

\bibitem[{{Chen} et~al.(2018){Chen}, {Lin}, {Alexandersen}, {Lehner}, {Wang},
  {Wang}, {Yoshida}, {Komiyama}, and {Miyazaki}}]{2018PASJ...70S..38C}
{Chen} YT, {Lin} HW, {Alexandersen} M, {Lehner} MJ, {Wang} SY, {Wang} JH,
  {Yoshida} F, {Komiyama} Y, {Miyazaki} S (2018) {Searching for moving objects
  in HSC-SSP: Pipeline and preliminary results}. Publications of the
  Astronomical Society of Japan 70:S38, \doi{10.1093/pasj/psx145},
  \eprint{1705.01722}

\bibitem[{Chipman et~al.(2010)Chipman, George, and McCulloch}]{Chipman_2010}
Chipman HA, George EI, McCulloch RE (2010) Bart: Bayesian additive regression
  trees. The Annals of Applied Statistics 4(1), \doi{10.1214/09-aoas285},
  \urlprefix\url{http://dx.doi.org/10.1214/09-AOAS285}

\bibitem[{{Chollet} and {others}(2018)}]{Chollet_2018}
{Chollet} F, {others} (2018) {Keras: The Python Deep Learning library}

\bibitem[{{Cincotta} and {Sim{\'o}}(2000)}]{cincotta_2000}
{Cincotta} PM, {Sim{\'o}} C (2000) {Simple tools to study global dynamics in
  non-axisymmetric galactic potentials - I}. Astronomy \& Astrophysics,
  Supplement 147:205--228, \doi{10.1051/aas:2000108}

\bibitem[{Cortes and Vapnik(2009)}]{Cortes_09}
Cortes C, Vapnik V (2009) Support-vector networks. Chem Biol Drug Des
  297:273--297, \doi{10.1007/%2FBF00994018}

\bibitem[{Cramer(2004)}]{CRAMER2004613}
Cramer J (2004) The early origins of the logit model. Studies in History and
  Philosophy of Science Part C: Studies in History and Philosophy of Biological
  and Biomedical Sciences 35(4):613--626,
  \doi{https://doi.org/10.1016/j.shpsc.2004.09.003},
  \urlprefix\url{https://www.sciencedirect.com/science/article/pii/S1369848604000676}

\bibitem[{{Dalpiaz et al.}(2021)}]{Dalpiaz_2021}
{Dalpiaz et al} (2021) Applied Statistics with R, STAT 420. University of
  Illinois at Urbana-Champaign,
  \url{https://daviddalpiaz.github.io/appliedstats/}

\bibitem[{{de Souza} et~al.(2021){de Souza}, {Krone-Martins}, {Carruba}, {de
  Cassia Domingos}, {Ishida}, {Alijbaae}, {Huaman Espinoza}, and
  {Barletta}}]{2021RNAAS...5..199D}
{de Souza} RS, {Krone-Martins} A, {Carruba} V, {de Cassia Domingos} R, {Ishida}
  EEO, {Alijbaae} S, {Huaman Espinoza} M, {Barletta} W (2021) {Probabilistic
  Modeling of Asteroid Diameters from Gaia DR2 Errors}. Research Notes of the
  American Astronomical Society 5(8):199, \doi{10.3847/2515-5172/ac205e},
  \eprint{2108.11814}

\bibitem[{{Dekany} et~al.(2020){Dekany}, {Smith}, {Riddle}, {Feeney}, {Porter},
  {Hale}, {Zolkower}, {Belicki}, {Kaye}, {Henning}, {Walters}, {Cromer},
  {Delacroix}, {Rodriguez}, {Reiley}, {Mao}, {Hover}, {Murphy}, {Burruss},
  {Baker}, {Kowalski}, {Reif}, {Mueller}, {Bellm}, {Graham}, and
  {Kulkarni}}]{2020PASP..132c8001D}
{Dekany} R, {Smith} RM, {Riddle} R, {Feeney} M, {Porter} M, {Hale} D,
  {Zolkower} J, {Belicki} J, {Kaye} S, {Henning} J, {Walters} R, {Cromer} J,
  {Delacroix} A, {Rodriguez} H, {Reiley} DJ, {Mao} P, {Hover} D, {Murphy} P,
  {Burruss} R, {Baker} J, {Kowalski} M, {Reif} K, {Mueller} P, {Bellm} E,
  {Graham} M, {Kulkarni} SR (2020) {The Zwicky Transient Facility: Observing
  System}. Publications of the Astronomical Society of the Pacific
  132(1009):038001, \doi{10.1088/1538-3873/ab4ca2}, \eprint{2008.04923}

\bibitem[{Dickey and Fuller(1979)}]{doi:10.1080/01621459.1979.10482531}
Dickey DA, Fuller WA (1979) Distribution of the estimators for autoregressive
  time series with a unit root. Journal of the American Statistical Association
  74(366a):427--431, \doi{10.1080/01621459.1979.10482531},
  \urlprefix\url{https://doi.org/10.1080/01621459.1979.10482531},
  \eprint{https://doi.org/10.1080/01621459.1979.10482531}

\bibitem[{{Duev} et~al.(2019){Duev}, {Mahabal}, {Ye}, {Tirumala}, {Belicki},
  {Dekany}, {Frederick}, {Graham}, {Laher}, {Masci}, {Prince}, {Riddle},
  {Rosnet}, and {Soumagnac}}]{2019MNRAS.486.4158D}
{Duev} DA, {Mahabal} A, {Ye} Q, {Tirumala} K, {Belicki} J, {Dekany} R,
  {Frederick} S, {Graham} MJ, {Laher} RR, {Masci} FJ, {Prince} TA, {Riddle} R,
  {Rosnet} P, {Soumagnac} MT (2019) {DeepStreaks: identifying fast-moving
  objects in the Zwicky Transient Facility data with deep learning}. \mnras
  486(3):4158--4165, \doi{10.1093/mnras/stz1096}, \eprint{1904.05920}

\bibitem[{{Duev} et~al.(2021){Duev}, {Bolin}, {Graham}, {Kelley}, {Mahabal},
  {Bellm}, {Coughlin}, {Dekany}, {Helou}, {Kulkarni}, {Masci}, {Prince},
  {Riddle}, {Soumagnac}, and {van der Walt}}]{2021AJ....161..218D}
{Duev} DA, {Bolin} BT, {Graham} MJ, {Kelley} MSP, {Mahabal} A, {Bellm} EC,
  {Coughlin} MW, {Dekany} R, {Helou} G, {Kulkarni} SR, {Masci} FJ, {Prince} TA,
  {Riddle} R, {Soumagnac} MT, {van der Walt} SJ (2021) {Tails: Chasing Comets
  with the Zwicky Transient Facility and Deep Learning}. \aj 161(5):218,
  \doi{10.3847/1538-3881/abea7b}, \eprint{2102.13352}

\bibitem[{{Erasmus} et~al.(2017){Erasmus}, {Mommert}, {Trilling}, {Sickafoose},
  {van Gend}, and {Hora}}]{2017AJ....154..162E}
{Erasmus} N, {Mommert} M, {Trilling} DE, {Sickafoose} AA, {van Gend} C, {Hora}
  JL (2017) {Characterization of Near-Earth Asteroids Using KMTNET-SAAO}. \aj
  154(4):162, \doi{10.3847/1538-3881/aa88be}, \eprint{1709.03305}

\bibitem[{{Erasmus} et~al.(2018){Erasmus}, {McNeill}, {Mommert}, {Trilling},
  {Sickafoose}, and {van Gend}}]{2018ApJS..237...19E}
{Erasmus} N, {McNeill} A, {Mommert} M, {Trilling} DE, {Sickafoose} AA, {van
  Gend} C (2018) {Taxonomy and Light-curve Data of 1000 Serendipitously
  Observed Main-belt Asteroids}. The Astrophysical Journal Supplement Series
  237(1):19, \doi{10.3847/1538-4365/aac38f}, \eprint{1805.04478}

\bibitem[{{Feigelson} et~al.(2018){Feigelson}, {Babu}, and
  {Caceres}}]{2018FrP.....6...80F}
{Feigelson} ED, {Babu} GJ, {Caceres} GA (2018) {Autoregressive Times Series
  Methods for Time Domain Astronomy}. Frontiers in Physics 6:80,
  \doi{10.3389/fphy.2018.00080}, \eprint{1901.08003}

\bibitem[{{Florios} et~al.(2018){Florios}, {Kontogiannis}, {Park}, {Guerra},
  {Benvenuto}, {Bloomfield}, and {Georgoulis}}]{2018SoPh..293...28F}
{Florios} K, {Kontogiannis} I, {Park} SH, {Guerra} JA, {Benvenuto} F,
  {Bloomfield} DS, {Georgoulis} MK (2018) {Forecasting Solar Flares Using
  Magnetogram-based Predictors and Machine Learning}. Solar Physics 293(2):28,
  \doi{10.1007/s11207-018-1250-4}, \eprint{1801.05744}

\bibitem[{{Fluke} and {Jacobs}(2020)}]{2020WDMKD..10.1349F}
{Fluke} CJ, {Jacobs} C (2020) {Surveying the reach and maturity of machine
  learning and artificial intelligence in astronomy}. WIREs Data Mining and
  Knowledge Discovery 10(2):e1349, \doi{10.1002/widm.1349}, \eprint{1912.02934}

\bibitem[{Freund and Schapire(1999)}]{Freund_99}
Freund Y, Schapire R (1999) Large margin classification using the perceptron
  algorithm. Machine Learning 37, \doi{10.1023/A:1007662407062}

\bibitem[{Freund and Schapire(1995)}]{Freund95adecision-theoretic}
Freund Y, Schapire RE (1995) A decision-theoretic generalization of on-line
  learning and an application to boosting

\bibitem[{{Gaia Collaboration} et~al.(2018){Gaia Collaboration}, {Spoto},
  {Tanga}, and et~al.}]{2018A&A...616A..13G}
{Gaia Collaboration}, {Spoto} F, {Tanga} P, et~al (2018) {Gaia Data Release 2.
  Observations of solar system objects}. \aap 616:A13,
  \doi{10.1051/0004-6361/201832900}, \eprint{1804.09379}

\bibitem[{Gowanlock et~al.(2021)Gowanlock, Kramer, Trilling, Butler, and
  Donnelly}]{Gowanlock2021FastPS}
Gowanlock MG, Kramer DA, Trilling DE, Butler NR, Donnelly B (2021) Fast period
  searches using the lomb-scargle algorithm on graphics processing units for
  large datasets and real-time applications. Astron Comput 36:100472

\bibitem[{Gudivada et~al.(2016)Gudivada, Irfan, Fathi, and
  Rao}]{GUDIVADA2016169}
Gudivada V, Irfan M, Fathi E, Rao D (2016) Chapter 5 - cognitive analytics:
  Going beyond big data analytics and machine learning. In: Gudivada VN,
  Raghavan VV, Govindaraju V, Rao C (eds) Cognitive Computing: Theory and
  Applications, Handbook of Statistics, vol~35, Elsevier, pp 169--205,
  \doi{https://doi.org/10.1016/bs.host.2016.07.010},
  \urlprefix\url{https://www.sciencedirect.com/science/article/pii/S0169716116300517}

\bibitem[{Hill et~al.(2020)Hill, Linero, and
  Murray}]{doi:10.1146/annurev-statistics-031219-041110}
Hill J, Linero A, Murray J (2020) Bayesian additive regression trees: A review
  and look forward. Annual Review of Statistics and Its Application
  7(1):251--278, \doi{10.1146/annurev-statistics-031219-041110},
  \urlprefix\url{https://doi.org/10.1146/annurev-statistics-031219-041110},
  \eprint{https://doi.org/10.1146/annurev-statistics-031219-041110}

\bibitem[{Ho(1995)}]{Ho95}
Ho TK (1995) Random decision forests. In: Proceedings of the Third
  International Conference on Document Analysis and Recognition (Volume 1) -
  Volume 1, IEEE Computer Society, M, ICDAR '95, pp 278--282

\bibitem[{Ho(1998)}]{Ho98}
Ho TK (1998) The random subspace method for constructing decision forests. IEEE
  Transactions on Pattern Analysis and Machine Intelligence 20(8):832--844,
  \doi{10.1109/34.709601}

\bibitem[{Jones et~al.(2015)Jones, Juri\'{c}, and Ivezi\'{c}}]{Lynne_2015}
Jones RL, Juri\'{c} M, Ivezi\'{c} v (2015) Asteroid discovery and
  characterization with the large synoptic survey telescope. Proceedings of the
  International Astronomical Union 10(S318):282–292,
  \doi{10.1017/s1743921315008510},
  \urlprefix\url{http://dx.doi.org/10.1017/S1743921315008510}

\bibitem[{{Li} et~al.(2022){Li}, {Li}, {Xia}, and
  {Georgakarakos}}]{2022MNRAS.511.2218L}
{Li} X, {Li} J, {Xia} ZJ, {Georgakarakos} N (2022) {Machine-learning prediction
  for mean motion resonance behaviour - The planar case}. \mnras
  511(2):2218--2228, \doi{10.1093/mnras/stac166}, \eprint{2201.06743}

\bibitem[{{Lin} et~al.(2018){Lin}, {Chen}, {Wang}, {Wang}, {Yoshida}, {Ip},
  {Miyazaki}, and {Terai}}]{2018PASJ...70S..39L}
{Lin} HW, {Chen} YT, {Wang} JH, {Wang} SY, {Yoshida} F, {Ip} WH, {Miyazaki} S,
  {Terai} T (2018) {Machine-learning-based real-bogus system for the HSC-SSP
  moving object detection pipeline}. Publications of the Astronomical Society
  of Japan 70:S39, \doi{10.1093/pasj/psx082}, \eprint{1704.06413}

\bibitem[{{Liu} et~al.(2021){Liu}, {Gong}, and {Li}}]{2021MNRAS.502.5362L}
{Liu} C, {Gong} S, {Li} J (2021) {Stability time-scale prediction for main-belt
  asteroids using neural networks}. \mnras 502(4):5362--5369,
  \doi{10.1093/mnras/stab080}

\bibitem[{Liu et~al.(2008)Liu, Ting, and Zhou}]{Liu_08}
Liu FT, Ting KM, Zhou ZH (2008) Isolation forest. In: 2008 Eighth IEEE
  International Conference on Data Mining, pp 413--422,
  \doi{10.1109/ICDM.2008.17}

\bibitem[{MacQueen(1967)}]{MacQueen1967}
MacQueen JB (1967) Some methods for classification and analysis of multivariate
  observations. In: Cam LML, Neyman J (eds) Proc. of the fifth Berkeley
  Symposium on Mathematical Statistics and Probability, University of
  California Press, vol~1, pp 281--297

\bibitem[{Mahabal et~al.(2019)Mahabal, Rebbapragada, Walters, Masci,
  Blagorodnova, van Roestel, Ye, Biswas, Burdge, Chang, Duev, Golkhou, Miller,
  Nordin, Ward, Adams, Bellm, Branton, Bue, Cannella, Connolly, Dekany, Feindt,
  Hung, Fortson, Frederick, Fremling, Gezari, Graham, Groom, Kasliwal,
  Kulkarni, Kupfer, Lin, Lintott, Lunnan, Parejko, Prince, Riddle, Rusholme,
  Saunders, Sedaghat, Shupe, Singer, Soumagnac, Szkody, Tachibana, Tirumala,
  van Velzen, and Wright}]{Mahabal_2019}
Mahabal A, Rebbapragada U, Walters R, Masci FJ, Blagorodnova N, van Roestel J,
  Ye QZ, Biswas R, Burdge K, Chang CK, Duev DA, Golkhou VZ, Miller AA, Nordin
  J, Ward C, Adams S, Bellm EC, Branton D, Bue B, Cannella C, Connolly A,
  Dekany R, Feindt U, Hung T, Fortson L, Frederick S, Fremling C, Gezari S,
  Graham M, Groom S, Kasliwal MM, Kulkarni S, Kupfer T, Lin HW, Lintott C,
  Lunnan R, Parejko J, Prince TA, Riddle R, Rusholme B, Saunders N, Sedaghat N,
  Shupe DL, Singer LP, Soumagnac MT, Szkody P, Tachibana Y, Tirumala K, van
  Velzen S, Wright D (2019) Machine learning for the zwicky transient facility.
  Publications of the Astronomical Society of the Pacific 131(997):038002,
  \doi{10.1088/1538-3873/aaf3fa},
  \urlprefix\url{https://doi.org/10.1088/1538-3873/aaf3fa}

\bibitem[{{Malhotra} et~al.(2015){Malhotra}, {Vig}, {Shroff}, and
  {Agarwal}}]{Pankaj_2015}
{Malhotra} P, {Vig} L, {Shroff} G, {Agarwal} P (2015) {Long Short Term Memory
  Networks for Anomaly Detection in Time Series}. In: European Symposium on
  Artificial Neural Networks, Computational Intelligence and Machine Learning,
  ESANN

\bibitem[{Mills(1999)}]{RePEc:elg:eebook:1506}
Mills TC (ed)  (1999) Economic Forecasting, 1-2, vol Two volume set. Edward
  Elgar Publishing,
  \urlprefix\url{https://EconPapers.repec.org/RePEc:elg:eebook:1506}

\bibitem[{{Mommert} et~al.(2016){Mommert}, {Trilling}, {Borth}, {Jedicke},
  {Butler}, {Reyes-Ruiz}, {Pichardo}, {Petersen}, {Axelrod}, and
  {Moskovitz}}]{2016AJ....151...98M}
{Mommert} M, {Trilling} DE, {Borth} D, {Jedicke} R, {Butler} N, {Reyes-Ruiz} M,
  {Pichardo} B, {Petersen} E, {Axelrod} T, {Moskovitz} N (2016) {First Results
  from the Rapid-response Spectrophotometric Characterization of Near-Earth
  Objects using UKIRT}. \aj 151(4):98, \doi{10.3847/0004-6256/151/4/98},
  \eprint{1602.06000}

\bibitem[{{Moschini} et~al.(2020){Moschini}, {Houssou}, {Bovay}, and
  {Robert-Nicoud}}]{2020arXiv200907578M}
{Moschini} G, {Houssou} R, {Bovay} J, {Robert-Nicoud} S (2020) {Anomaly and
  Fraud Detection in Credit Card Transactions Using the ARIMA Model}. arXiv
  e-prints arXiv:2009.07578, \eprint{2009.07578}

\bibitem[{{Pearson}(1895)}]{1895RSPS...58..240P}
{Pearson} K (1895) {Note on Regression and Inheritance in the Case of Two
  Parents}. Proceedings of the Royal Society of London Series I 58:240--242

\bibitem[{{Pedregosa} et~al.(2012){Pedregosa}, {Varoquaux}, {Gramfort},
  {Michel}, {Thirion}, {Grisel}, {Blondel}, {M{\"u}ller}, {Nothman}, {Louppe},
  {Prettenhofer}, {Weiss}, {Dubourg}, {Vanderplas}, {Passos}, {Cournapeau},
  {Brucher}, {Perrot}, and {Duchesnay}}]{2012arXiv1201.0490P}
{Pedregosa} F, {Varoquaux} G, {Gramfort} A, {Michel} V, {Thirion} B, {Grisel}
  O, {Blondel} M, {M{\"u}ller} A, {Nothman} J, {Louppe} G, {Prettenhofer} P,
  {Weiss} R, {Dubourg} V, {Vanderplas} J, {Passos} A, {Cournapeau} D, {Brucher}
  M, {Perrot} M, {Duchesnay} {\'E} (2012) {Scikit-learn: Machine Learning in
  Python}. arXiv e-prints arXiv:1201.0490, \eprint{1201.0490}

\bibitem[{{Penttil{\"a}} et~al.(2021){Penttil{\"a}}, {Hietala}, and
  {Muinonen}}]{Pentilla_2021}
{Penttil{\"a}} A, {Hietala} H, {Muinonen} K (2021) {Asteroid spectral taxonomy
  using neural networks}. A\&A 649:A46, \doi{10.1051/0004-6361/202038545}

\bibitem[{Penttilä et~al.(2022)Penttilä, Fedorets, and
  Muinonen}]{10.3389/fspas.2022.816268}
Penttilä A, Fedorets G, Muinonen K (2022) Taxonomy of asteroids from the
  legacy survey of space and time using neural networks. Frontiers in Astronomy
  and Space Sciences 9, \doi{10.3389/fspas.2022.816268},
  \urlprefix\url{https://www.frontiersin.org/article/10.3389/fspas.2022.816268}

\bibitem[{{Pesenson} et~al.(2010){Pesenson}, {Pesenson}, and
  {McCollum}}]{2010AdAst2010E..58P}
{Pesenson} MZ, {Pesenson} IZ, {McCollum} B (2010) {The Data Big Bang and the
  Expanding Digital Universe: High-Dimensional, Complex and Massive Data Sets
  in an Inflationary Epoch}. Advances in Astronomy 2010:350891,
  \doi{10.1155/2010/350891}, \eprint{1003.0879}

\bibitem[{Piryonesi and El-Diraby(2020)}]{Piryonesi_20}
Piryonesi SM, El-Diraby T (2020) Data analytics in asset management:
  Cost-effective prediction of the pavement condition. Journal of
  Infrastructure Systems 26, \doi{10.1061/(ASCE)IS.1943-555X.0000512}

\bibitem[{{Pravec} et~al.(2010){Pravec}, {Vokrouhlick{\'y}}, {Polishook},
  {Scheeres}, {Harris}, {Gal{\'a}d}, {Vaduvescu}, {Pozo}, {Barr}, {Longa},
  {Vachier}, {Colas}, {Pray}, {Pollock}, {Reichart}, {Ivarsen}, {Haislip},
  {Lacluyze}, {Ku{\v{s}}nir{\'a}k}, {Henych}, {Marchis}, {Macomber},
  {Jacobson}, {Krugly}, {Sergeev}, and {Leroy}}]{2010Natur.466.1085P}
{Pravec} P, {Vokrouhlick{\'y}} D, {Polishook} D, {Scheeres} DJ, {Harris} AW,
  {Gal{\'a}d} A, {Vaduvescu} O, {Pozo} F, {Barr} A, {Longa} P, {Vachier} F,
  {Colas} F, {Pray} DP, {Pollock} J, {Reichart} D, {Ivarsen} K, {Haislip} J,
  {Lacluyze} A, {Ku{\v{s}}nir{\'a}k} P, {Henych} T, {Marchis} F, {Macomber} B,
  {Jacobson} SA, {Krugly} YN, {Sergeev} AV, {Leroy} A (2010) {Formation of
  asteroid pairs by rotational fission}. Nature 466(7310):1085--1088,
  \doi{10.1038/nature09315}, \eprint{1009.2770}

\bibitem[{{Pugliatti} and Topputo(2020)}]{pugliatti_2020}
{Pugliatti} M, Topputo F (2020) Small-body shape recognition with convolutional
  neural network and comparison with explicit features based methods. AAS/AIAA
  Astrodynamics Specialist Conference pp 1--20

\bibitem[{{R Core Team}(2013)}]{R_ref}
{R Core Team} (2013) R: A Language and Environment for Statistical Computing. R
  Foundation for Statistical Computing, Vienna, Austria,
  \urlprefix\url{http://www.R-project.org/}

\bibitem[{Rosenblatt(1963)}]{Rosenblatt1963PRINCIPLESON}
Rosenblatt F (1963) Principles of neurodynamics. perceptrons and the theory of
  brain mechanisms. American Journal of Psychology 76:705

\bibitem[{van Rossum(1995)}]{CS-R9526}
van Rossum G (1995) Python tutorial. Tech. Rep. CS-R9526, Centrum voor Wiskunde
  en Informatica (CWI), Amsterdam

\bibitem[{Russell and Norvig(2010)}]{russel2010}
Russell S, Norvig P (2010) Artificial Intelligence: A Modern Approach, 3rd edn.
  Prentice Hall

\bibitem[{{Sanchez} and {Prado}(2017)}]{sanchez_2017}
{Sanchez} DM, {Prado} AFBA (2017) {On the Use of Mean Motion Resonances to
  Explore the Haumea System}. AAS/AIAA Astrodynamics Specialist Conference
  162:1507--1524

\bibitem[{{Sanchez} and {Prado}(2019)}]{sanchez_2019}
{Sanchez} DM, {Prado} AFBA (2019) {Searching for Less-Disturbed Orbital Regions
  Around the Near-Earth Asteroid 2001 SN263}. Journal of Spacecraft and Rockets
  56(6):1775--1785, \doi{10.2514/1.A34402}

\bibitem[{Sander et~al.(1998)Sander, Ester, Kriegel, and Xiaowei}]{DBSCAN_1998}
Sander J, Ester M, Kriegel H, Xiaowei X (1998) Density-based clustering in
  spatial databases: The algorithm gdbscan and its applications. Data Mining
  and Knowledge Discovery 2:169--194

\bibitem[{Seabold and Perktold(2010)}]{seabold2010statsmodels}
Seabold S, Perktold J (2010) statsmodels: Econometric and statistical modeling
  with python. In: 9th Python in Science Conference

\bibitem[{{Smirnov} and {Markov}(2017)}]{2017MNRAS.469.2024S}
{Smirnov} EA, {Markov} AB (2017) {Identification of asteroids trapped inside
  three-body mean motion resonances: a machine-learning approach}. MNRAS
  469(2):2024--2031, \doi{10.1093/mnras/stx999}

\bibitem[{{Smullen} and {Volk}(2020)}]{2020MNRAS.497.1391S}
{Smullen} RA, {Volk} K (2020) {Machine learning classification of Kuiper belt
  populations}. MNRAS 497(2):1391--1403, \doi{10.1093/mnras/staa1935},
  \eprint{2007.03720}

\bibitem[{Strigl et~al.(2010)Strigl, Kofler, and Podlipnig}]{5452452}
Strigl D, Kofler K, Podlipnig S (2010) Performance and scalability of gpu-based
  convolutional neural networks. In: 18th Euromicro International Conference on
  Parallel, Distributed and Network-Based Processing (PDP 2010), IEEE Computer
  Society, Los Alamitos, CA, USA, \doi{10.1109/PDP.2010.43},
  \urlprefix\url{https://doi.ieeecomputersociety.org/10.1109/PDP.2010.43}

\bibitem[{Vaswani et~al.(2017)Vaswani, Shazeer, Parmar, Uszkoreit, Jones,
  Gomez, Kaiser, and Polosukhin}]{vaswani2017attention}
Vaswani A, Shazeer N, Parmar N, Uszkoreit J, Jones L, Gomez AN, Kaiser L,
  Polosukhin I (2017) Attention is all you need. \eprint{1706.03762}

\bibitem[{Wang(2001)}]{Wang_2001}
Wang D (2001) Unsupervised learning: Foundations of neural computation. AI
  Magazine 22(2):101, \doi{10.1609/aimag.v22i2.1565},
  \urlprefix\url{https://ojs.aaai.org/index.php/aimagazine/article/view/1565}

\bibitem[{{Yazdanbakhsh} et~al.(2021){Yazdanbakhsh}, {Seshadri}, {Akin},
  {Laudon}, and {Narayanaswami}}]{2021arXiv210210423Y}
{Yazdanbakhsh} A, {Seshadri} K, {Akin} B, {Laudon} J, {Narayanaswami} R (2021)
  {An Evaluation of Edge TPU Accelerators for Convolutional Neural Networks}.
  arXiv e-prints arXiv:2102.10423, \eprint{2102.10423}

\bibitem[{{Zhang} et~al.(2018){Zhang}, {Zhang}, and
  {Zhao}}]{2018AJ....155..108Z}
{Zhang} J, {Zhang} Y, {Zhao} Y (2018) {Imbalanced Learning for RR Lyrae Stars
  Based on SDSS and GALEX Databases}. \aj 155(3):108,
  \doi{10.3847/1538-3881/aaa5b1}

\end{thebibliography}

\end{document}